 \documentclass[11pt]{article}

 \pdfoutput=1   

\usepackage{graphicx}
\usepackage{amsmath}

\usepackage{enumitem}

\usepackage[francais,english]{babel}

\usepackage{float}
\usepackage{ifthen}
\usepackage{color}

\usepackage{hyperref}

\begin{document}


\title{On the concept of exergy and available enthalpy: \\
       Application to atmospheric energetics.}

\author{by Pascal Marquet. {\it M\'et\'eo-France.}}


\date{\today}
\maketitle


\vspace*{-10mm}

\begin{center}
{\em Copy of a paper submitted in April 1990 to the \underline{Quarterly Journal of the Royal Meteorological Society}.} \\
{\em Published in Volume 117, Issue 499, pages 449-475, April 1991}: \\
\url{http://onlinelibrary.wiley.com/doi/10.1002/qj.49711749903/abstract} \\
Several comments and corrections are added in the main text and in footnotes. \\
Replies are given in Appendix~C about the note of Dutton (1992):\\
\url{http://onlinelibrary.wiley.com/doi/10.1002/qj.49711850309/abstract} \\
{\em \underline{Corresponding address}: pascal.marquet@meteo.fr}
\end{center}
\vspace{1mm}


\vspace*{-2mm}

\begin{abstract}
The available enthalpy is an early form of the modern thermodynamic concept of exergy, which is the generic name for the amount of work obtainable when some matter is brought to a state of equilibrium with its surroundings by means of reversible processes.

It is shown in this paper that a study of the hydrodynamic properties of available enthalpy leads to a generalization of the global meteorological available energies previously introduced by Lorenz, Dutton and Pearce. 
A local energy cycle is derived without approximation. Moreover, static instabilities or topography do not prevent this theory from having practical applications. 
The concept of available enthalpy is also presented in terms of the potential change in total entropy. Using the hydrostatic assumption, limited-area energetics is then rigorously defined, including new boundary fluxes and new energy components. 
This innovative approach is especially suitable for the study of energy conversions between isobaric layers of an open limited atmospheric domain.

Numerical evaluations of various energy components are presented for a hemispheric field of zonal-average temperature. 
It is further shown that this new energetic scheme realizes a hierarchical partition of the components so that the smallest of those available enthalpy reservoirs are almost of the same magnitude as the kinetic energy. 
This is actually the fundamental property that induced Margules to define the primary concept of available kinetic energy in meteorology.
\end{abstract}


 \section{INTRODUCTION.} 
\label{section_INTRO}

The concept of available kinetic energy was introduced in atmospheric energetics by Margules (1905) to explain the generation of strong winds in storms. 
He defined it as the maximum kinetic energy that an isolated and closed mass of air at rest can generate when undergoing adiabatic changes in its thermodynamic structure. 

Lorenz (1955, 1967) developed these ideas by applying them to the general circulation of the atmosphere.
His aim was to diagnose the energy sources, sinks and conversion terms in meteorology to understand the maintenance of atmospheric motions despite dissipation by friction.
He coined the term ``Available Potential Energy'' (APE) and defined it as being the fraction of total potential energy (TPE) -- the sum of internal and potential energies over the whole atmosphere -- that could potentially be transformed into kinetic energy through isentropic processes. 

The APE is the difference between the TPE of the real state of the atmosphere and the TPE of an associated reference state. 
This reference state is determined from the real one by an adiabatic redistribution of the mass until all the atmosphere reaches horizontal and statically stable stratification (the mass between two isentropes being conserved). 
This reference state corresponds to a minimum TPE and to a maximum kinetic energy among all possible adiabatic redistributions, but it may not be dynamically reached since only frictionally modified inertial flows are allowed by such a stratification of thermodynamic fields. 
Therefore, the definition of a relevant reference state is particularly crucial to unravel the mechanism which maintains the atmospheric motions.

Van Mieghem (1956) used a variational approach to show that different reference states can indeed be used to define an APE. Dutton and Johnson (1967) argued that this type of treatment cannot explain why particular modes of atmospheric circulation are observed, since the underlying assumption that the atmosphere tries to reach these reference states has never been established.

Starting from these conclusions, Pearce (1978) made a successful attempt to give up Lorenz's reference state. 
He defined the available energy only by specifying a set of properties that energy sources, sinks or conversion terms should possess: 
\begin{itemize}[label=,leftmargin=3mm,parsep=0cm,itemsep=0.1cm,topsep=0cm,rightmargin=2mm]
\vspace*{-1mm}
\item  (i) the conversion term must be opposite to the usual term in the kinetic energy equation, $-\: \omega\: \alpha$ (see list of symbols in Appendix~A); 
\item  (ii) spatial variations of the distributions of heat sources and sinks must generate available energy whereas uniform distributions rather tend to increase the huge unavailable energy reservoir.
\end{itemize}

The approximate APE as defined by Lorenz depends on the ratio of two terms associated with two distinct physical phenomena. 
The first, which appears in the numerator, is the isobaric variance of temperature and corresponds to a baroclinicity effect.
The second, which appears in the denominator, is a pressure-dependent stability parameter (but it is assumed to be time-independent). 
Pearce (1978) obtained a sum of two available energy components which separates and emphasizes these two phenomena. 
He defined the first energy reservoir as the baroclinicity component, which only depends on the isobaric variance of the temperature, and the second energy reservoir as the static stability component, which is associated with the deviations from an isothermal atmosphere of the vertical profile of the isobaric average of the temperature. 

It is well known that the approximate form of the APE, as defined by Lorenz, turns out to have unfortunate properties as shown by Dutton and Johnson (1967), among others. 
But the exact form of the APE has not been widely applied in energetics studies. 
Dutton (1973, 1976) introduced the global concept of static entropic energy ($T_0\:\Sigma$).
The underlying physical notion depends on the Gibbs equilibrium theorem which tells us that the state of maximum entropy (among those with the same total mass $M$ and total energy $H$) is the more stable state: it is the associated reference isothermal state which is thus thermodynamically stable (one can also recognize the isothermal atmosphere of Pearce). 
Then Dutton (1973, 1976) introduced several theorems related to the natural time-trend of the atmosphere, taking advantage of the effect of the second law of thermodynamics which characterizes this natural trend under certain assumptions.

Local physical interpretations of the available energy have not been obtained using the formulations of Lorenz, Dutton or Pearce. 
Indeed, energy studies over limited area (Oort 1964; Muench 1965; Brenan and Vincent 1980; Michaelides 1987) have used unsuitable mathematical tools. All these studies have dealt with energy integrals. 
The use of integration by parts with boundary conditions at infinity makes it impossible to return to the local definition of an energy cycle. In these atmospheric studies, certain equations require such boundary conditions. 
These equations consist of the values of the integrals of internal and gravitational potential energies which only become proportional for a vertically infinite atmosphere, the sum of these two integrals being then the total enthalpy of the whole atmosphere. 
Moreover, despite improvements made by Smith {\it et al.\/} (1977) and Boer (1989) concerning limited-area or isobaric coordinate problems, the definition of available energy using a reference state (following Lorenz's method) assumes an adiabatic redistribution of the mass solely restricted to the prescribed limited area.
Therefore, these reference states are not intrinsically defined. 

A thorough revision of the global concept of available energy is thus necessary in order to find a local counterpart (local budgets of Dutton's entropic energy have been derived by Pichler (1977) and Karlsson (1990) but the present author was unaware of this fact when this paper was first submitted for publication).

Like Lorenz and Pearce, Dutton dealt with global integrals over the whole atmosphere.
However, the static entropy energy $T_0\:\Sigma$ introduced by Dutton is the first application of exergy-like theory to atmospheric studies.
Indeed $T_0\:\Sigma$ is the global integral of the primary available energy as defined by most founders of thermodynamics (e.g. Kelvin, Maxwell, Gibbs) and the respective motionless associated reference states are the same.\footnote{\color{blue} A note by Dutton (1992) has been published soon after my paper. Dutton was one of my Referees. The other one was R. P. Pearce. 
Several comments about the note of Dutton are given in Apendix~C.}

The purpose of this paper is to re-examine the concept of available energy in order to define it locally. 
A local algebraic expression for available energy would allow studies of energy conversions within open domains (cyclones, baroclinic waves), all boundary fluxes being taken into account. 
Moreover, energetics of each particular level (or for a given layer) could then be performed to investigate troposphere-stratosphere energy exchanges, for instance.

In this paper a new concept of potential change in total entropy is introduced. 
It is closely connected with the exergy theory but it does not use the concept of an associated
reference state for the atmosphere. 
First, a historical background of the thermodynamic concept of exergy is set out in Section~\ref{section_HIST_BACK}. 
The meteorological specific available enthalpy is then defined in Section~\ref{section_Definitions}, together with the equations governing the specific energy components usable for atmospheric energetics. 
A local energy cycle is described in Section~\ref{section_physical_properties} and some basic local and global properties of conservation are derived. 
Two arbitrary constants which arise from the mathematical introduction of the available enthalpy can then be defined when comparing the global energetic behaviour of available enthalpy with some of Lorenz's fundamental results that it is desirable to keep.
The link with the previous theory of Pearce is established in Section~\ref{section_hydro_energetics} when considering the energy cycle for isobaric average energy components under the hydrostatic assumption. 
Section~\ref{section_summary} concerns a numerical evaluation of the various energy components compared with those using Lorenz's and Pearce's definitions. 
A list of symbols is given in Appendix~A.
A (corrected and augmented) review of Exergy formulae is given in Appendix~B.
New replies to comments of this paper published in Dutton (1992) are given in Appendix~C.

 \section{HISTORICAL BACKGROUND.} 
\label{section_HIST_BACK}

In the nineteenth century W. Thomson (1849\footnote{\color{blue}  
This entry was not described  in Marquet (1991)}, 
1853, 1879) (Lord Kelvin), Gibbs (1873b), Tait (1879) and Maxwell (1889), initiated the concept of ``motivity'' or ``available energy''.
They referred either to the maximum work of any kind (the gross work) that a system can produce for a prescribed outer temperature, or to the maximum work available to the outer medium (the shaft work).
The ``shaft work'' is obtained by disregarding the work like that due to forces deriving from a potential (gravitation) or from expansion (generated by changes in the volume of the system against the outer pressure force).
At about the same time, Gouy (1889) proposed the same theory.
Stodola (1898) also formulated the same law, being unaware of the previous studies.
A valuable historical review of the thermodynamic concepts of availability is presented by Haywood (1974).

Since then, numerous studies have defined analogous quantities with various names such as available energy (``\'energie utilisable'', Darrieus 1931), ``coenthalpy'' (``coenthalpie'', Borel 1987), ``physical exergy'' (Keenan 1951), ``available enthalpy'' (``enthalpie utilisable'' , Martinot-Lagarde, 1971) or availability function for steady flow. 
All these names refer either to $H - T_0 \: S$ or to $(H - H_0) - T_0 \: (S - S_0)$; (see Appendix~B).

The theory of exergy has recently been developed to deal with more general processes: open systems and non-steady flows (Borel 1987). The term ``exergy'' was coined by Rant (1956). 
The name ``available energy'' is now well established in meteorology and the choice of the name ``available enthalpy'' seems suitable.

The available enthalpy is generally defined in thermodynamics as the state function $A_h = (H - H_0) - T_0 \: (S - S_0)$ which exactly corresponds to the physical exergy for steady flow. 
It is an extrinsic state function: a joint property of a fluid flowing through a device in steady flow (enthalpy $H$ and entropy $S$) and of the outer medium which surrounds this device and is assumed to act as a heat reservoir at the constant temperature $T_0$. 
But even if $H - T_0 \: S$ is similar in form to the Gibbs function $G = H - T\:S$ (also called free enthalpy) except for this outer temperature $T_0$, the physical properties of these quantities are completely different.

Since no thermostat or shaft work can be defined for the atmosphere, which is in unsteady flow, it might seem that the concept of exergy cannot be applied in meteorology.\footnote{\label{footnote_Keenan}\color{blue} It is explicitly mentioned in Keenan (1951, p.183) that: ``{\it Virtually all problems which can be treated adequately by the methods of thermodynamics are terrestrial: that is, they relate to the behaviour of systems which are surrounded by an essentially infinite atmosphere. A \underline{major exception} to this latter generalization is found in the subject of \underline{meteorology} wherein the system under consideration is the atmosphere itself \/}''.
Differently, it is shown in Marquet (1991) that the concept of available enthalpy can be used for meteorological purpose too.
The main issue to be solved concerns the  definition of the the constant temperature $T_r$ and pressure $p_r$.}
However, an appropriate way to apply exergy theory to atmospheric energetics is to investigate the physical properties of the mathematical concept of available enthalpy $A_h$, independently on the way it is justified.

If $E_i$ is the total internal energy of the atmosphere and if the subscript '$0$' denotes a certain equilibrium state, the ``static entropic energy'' of Dutton (1973, 1976) can be rewritten as $T_0\:\Sigma = {E_i - (E_i)_0} - T_0 \:  (S - S_0)$ which is the form of the available energy as defined by Maxwell, for instance.

A generalized form of $T_0\:\Sigma$ is defined  by Livezey and Dutton (1976, see Eq.30, p.144) for determining the available energy of ocean.
It can be written ${E_i- (E_i)_0}+ p_0 \: (V- V_0) - T_0\: (S - S_0)$, which is called ``non-flow exergy'' by Haywood (1974).
$V$  and $V_0$ are the actual and reference finite volumes of the ocean\footnote{\color{blue} The term ``infinite'' used in Marquet (1991) was not correct.}.
Like Evans (1980), who studied this latter problem, Livezey and Dutton (1976, Eq. (25), p. 150) introduced a concept similar to what is called ``non-flow essergy'' by Haywood (1974). 
Therefore, since Dutton has obtained numerous global results with such exergy-like quantities defined for either atmosphere or ocean, it is not surprising that the local exergy-like function introduced hereafter can be relevant to the definition of a local energetics of the atmosphere.\footnote{\color{blue} The publication of my paper in 1991 has inspired the work of Kucharski (1997), suggested by A. Thorpe at the University of Reading.}

 \section{DEFINITIONS.} 
\label{section_Definitions}

      \subsection{The specific available enthalpy.} 
        \label{section_Ah}

The atmospheric fluid is assumed to be an ideal gas undergoing quasi-static thermodynamic processes (slow evolutions of the thermodynamic state). 
The moisture factor will not appear explicitly in the equation of state, although the heating rate associated with the saturation processes and changes of phase will be retained as the diabatic heating term (common meteorological practice). 
The pressure $p$, the absolute temperature $T$ and the density $\rho$ are linked through the equation of state $p = \rho \: R \: T$.

Introducing the undefined constant temperature $T_r$, the specific (i.e. expressed per unit mass) internal energy and enthalpy are
\begin{equation}
e_i \; = \;  (e_i)_r \: + \: c_v \: (T - T_r)
\nonumber
\end{equation}
and
\begin{equation}
h \; = \;  h_r \: + \: c_p \: (T - T_r) \: ,
\nonumber
\end{equation}
where $(e_i)_r$ and $h_r$  are the specific values related to the temperature $T_r$
(see notations in Appendix~A). 
For this ideal gas, the specific entropy can be defined as
\begin{equation}
s(\theta) \; = \;  s_r \: + \: c_p \: \ln(\theta / \theta_r) \: ,
\label{eq_1}
\end{equation}
where $\theta$ is the potential temperature and where
\begin{equation}
\theta_r \; = \; T_r \: (p_{00}/ p_r)^\kappa \: .
\nonumber
\end{equation}
$s_r$ is the uniform specific reference entropy related to the pressure $p_r$ and temperature
$T_r$ which are two undefined constants at this stage of the theory. 
Applying the first law of thermodynamics (associated local state assumption), the total derivative of $s(\theta)$ is
\begin{equation}
\frac{ds}{dt}
\; = \;  
\frac{c_p}{\theta}
\;
\frac{d\theta}{dt}
\; = \;  
\frac{Q}{T}
\: ,
\label{eq_2}
\end{equation}
where $Q$ is the specific heating rate due to radiation, conduction and latent heat release (neglecting molecular kinetic energy dissipation). 

Let us then define the specific available enthalpy mathematically as
\begin{equation}
\boxed{\;
a_h 
\; \equiv \;  
(h - h_r) \: - \: T_r \: (h - h_r) 
\; }
\; = \; 
(h - T_r \: s) \: - \: (h_r - T_r \: s_r) 
\:  .
\label{eq_3}
\end{equation}
Substituting $s - s_r$ from (\ref{eq_1}) into (\ref{eq_3})  and then expanding $\theta$ and $\theta_r$ leads to an expression for $a_h$ in terms of $T$ and $p$ ($T_r$ and $p_r$ being two constants):
\begin{equation}
a_h \:(T, p)
\; \equiv \;  
c_p  \: \left( T - T_r \right)
\: - \: 
c_p  \: T_r \: \ln \left( \frac{T}{T_r} \right)
\: + \: 
R \: T_r \: \ln \left( \frac{p}{p_r} \right)
\: ,
\nonumber
\end{equation}
which can be rewritten as
\begin{equation}
\boxed{ \;
a_h \:(T, p)
\; \equiv \;  
a_T \:(T)
\; + \; 
a_p \:( p) 
\;}
\: ,
\label{eq_4}
\end{equation}
where the specific available enthalpy is thus the sum of a positive component $a_T$ which
depends only on temperature:
\begin{equation}
\boxed{\;
a_T \:(T)
\; = \; 
c_p \: T_r \: F( X) 
\;}
\: , 
\; \; \; 
\mbox{where}
\; \; \; 
X \: = \: (T - T_r)/T_r \: = \: T /T_r - 1
\label{eq_5}
\end{equation}
\begin{equation}
\mbox{and for}
\; \; 
X>-1:
\; 
\boxed{\;
{\cal F}(X) 
\; = \;  X - \ln(1+X)
\;}
\; = \; \int^{X}_{0} \frac{x}{1+x} \: dx
\; \;
\Rightarrow
{\cal F}(X) \geq 0
\: ,
\label{eq_6}
\end{equation}
and of a component $a_p$ which depends only on pressure:
\begin{equation}
\boxed{\;
a_p(p)
\; = \;  
R \: T_r \: \ln(p/p_r)
\;}
\: .
\label{eq_7}
\end{equation}
Since $X > -1$ simply means $T> 0$, ${\cal F}(X)$ is always defined. 
It is a positive quantity, equal to zero only for $X= 0$, and thus for 
$T=T_r$.\footnote{\label{footnote_ap_Margules}{\color{blue} 
A criticism often reported is that $a_p(p)$ is not a positive quantity.
However, it is worth noting that $R \: T_r \: \ln(p/p_r)$}
{\color{blue} can be rewritten in terms of the derivative of a function ${\cal H}$ defined by ${\cal H}(X)=(1+X)\:\ln(1+X) - X$ and $X = p/p_r -1$, leading to $a_p = R \: T_r \: p_r \: d/dp[{\cal H}(X)]$.
This function ${\cal H}(X)$ is defined for $p>0$, i.e. for $X>-1$ as for ${\cal F}(X)$.
The leading approximation of ${\cal H}(X)$ for small $|X|$ is $X^2/2$ as for ${\cal F}(X)$.
It is easy to demonstrate that ${\cal H}(X)$ is positive, equal to zero only for $p=p_r$ and equal to $1$ for $p=0$. 
Accordingly, the global integral of $a_p$ (i.e. $A_p$) is roughly proportional to ${\cal H}(p_s/p_r-1)\approx (p_s-p_r)^2/(2\: p_r^2)$.
Moreover, accurate computations show that $A_p$ depends on the horizontal variance of surface pressure $\overline{(p_s-\overline{p_s})^2/ (\overline{p_s})^2}$.
This result was already derived in Margules (1901), in the paper dealing with ``The mechanical equivalent of any given distribution of atmospheric pressure, and the maintenance of a given difference in pressure''. 
This paper was published some years before the more famous one about the ``energy of storms'' (1903-05).}
}

      \subsection{Energy component equations.} 
        \label{section_energy_equations}

The energy equations to be derived in this section are the total derivatives of various energy components related to atmospheric energetics. 
To begin with $a_h$, the thermodynamic equation (\ref{eq_2})  can first be put in the form
\begin{equation}
\frac{dh}{dt}
\; = \;  
c_p \;
\frac{dT}{dt}
\; = \;  
\frac{R}{p}
\; \omega \: T
\; + \;  Q
\: ,
\label{eq_8}
\end{equation}
where $\omega = dp/dt$.
Equation (\ref{eq_8}) is the enthalpy equation.
Since $p_r$ and $T_r$ are two constants, the material derivative of $a_h$ can then be expressed as follows using  (\ref{eq_2}),  (\ref{eq_3}) and  (\ref{eq_8}):
\begin{equation}
\frac{d a_h}{dt}
\; = \;  
\frac{R}{p}
\; \omega \: T
\; + \; 
\left(  1 - \frac{T_r}{T} \right)
 Q
\: .
\label{eq_9}
\end{equation}
This available energy equation (\ref{eq_9}) also splits into the total derivatives of the two energy components $a_T$ and $a_p$
\begin{equation}
\frac{d a_T}{dt}
\; = \;  
- \: \frac{R}{p}
\; \omega \: T_r
\; + \; 
\frac{R}{p}
\; \omega \: T
\; + \; 
\left(  1 - \frac{T_r}{T} \right)
 Q
\; = \; 
\left(  1 - \frac{T_r}{T} \right)
\frac{d h}{dt}
\: ,
\label{eq_10}
\end{equation}
\begin{equation}
\frac{d a_p}{dt}
\; = \;  
+ \: \frac{R}{p}
\; \omega \: T_r
\: .
\label{eq_11}
\end{equation}

Taking the scalar product of the three-dimensional wind vector with the usual
momentum equation leads to the equation for $e_K = \vec{v} . \vec{v}/2$ ($e_K$ is the specific kinetic
energy) :
\begin{equation}
\frac{d e_K}{dt}
\; = \;  
- \: \frac{1}{\rho} \;
\vec{v} \: . \overrightarrow{\nabla}(p)
\: - \:
g \: w
\: + \:
\vec{v} \: . \:\vec{F} 
\: .
\label{eq_12}
\end{equation}

Finally, the fourth and last energy component is the specific gravitational potential
energy $e_G = \phi = g \: z$. 
The gravitational potential energy equation is thus
\begin{equation}
\frac{d e_G}{dt}
\; = \;  
  + \:
g \: w
\: .
\label{eq_13}
\end{equation}

 \section{PHYSICAL PROPERTIES.} 
\label{section_physical_properties}

In order to determine the unknown uniform constants $p_r$, and $T_r$, in this section the physical properties of the new components $a_T$ and $a_p$ are compared with the results of Lorenz (1955, 1967) and Pearce (1978).

      \subsection{Local energy cycle.} 
        \label{section_local_energy_cycle}

In the present study an energy cycle refers to a system of equations which consist of total derivatives of various energy components with respect to time, and to ``conversion terms'': expressions of opposite sign in two distinct equations involving in situ processes and having as far as possible a clear physical basis (Johnson and Downey 1982). 
The remaining terms are interpreted as sinks or sources. 
The local energy cycle with the set of energy components $\{ e_G$, $e_K$; $a_T$, $a_p\}$ is easily obtained from the energy equations (\ref{eq_10})  to (\ref{eq_13}) , expanding the total derivative $\omega$ in the equation for $a_T$ to find the conversion term $- C_{(h,K)}$:
\begin{equation}
\left.
\begin{aligned}
\frac{d e_G}{dt} & \; = \;  - \: C_{(G,K)} \\
\frac{d e_K}{dt} & \; = \;  + \: C_{(G,K)}      \;  + \: C_{(h,K)}       \;  - \:  D_K  \; \;  \\
\frac{d a_T}{dt} & \; = \;  + \: C_{(p,T)}   \: \;  -   \: C_{(h,K)}   \: \;  + \:  G_h \; \; \\
\frac{d a_p}{dt} & \; = \;  - \: C_{(p,T)} 
\end{aligned}
\right\}
\label{eq_14}
\end{equation}
where
\begin{equation}
 C_{(p,T)} \: = \:  
- \: \frac{R}{p} \; T_r \; \omega
   \; , \; \; \;
 C_{(h,K)} \: = \:   
- \: \frac{1}{\rho} \;
\vec{v} \: . \overrightarrow{\nabla}(p)
   \; , \; \; \;
 C_{(G,K)} \: = \:  - \: g \: w
\: 
\label{eq_15}
\end{equation}
and
\begin{equation}
 D_K \: = \: - \:  \vec{v} \: .  \: \vec{F} 
   \; , \; \; \;
 G_h \: = \:   
   \frac{1}{\rho} \;  \frac{\partial p}{\partial t} 
    \: + \:   \left(  1 - \frac{T_r}{T} \right) \: Q
\: .
\label{eq_16}
\end{equation}
The use of the total derivatives ensures a local physical meaning not only to the conversion terms (\ref{eq_15})  and source/sink terms  (\ref{eq_16}), but also to the temperature component $a_T$  and the pressure component $a_p$. 
The local cycle  (\ref{eq_14})  is an energetic balance obtained when following the motion of an atmospheric parcel.

A conversion term $C(\alpha, \beta)$  denotes a transfer from $\alpha$-energy to $\beta$-energy. 
The terms $C_{(G,K)}$ and $C_{(h,K)}$, the first term of $G_h$ and the dissipation term $D_K$ are already defined in the more ordinary enthalpy cycle $\{ \: e_G, e_K ; h \: \}$. 
In contrast, the conversion term $C_{(p,T)}$ for a transfer between $a_p$  and $a_T$ is inherent in the choice of the available enthalpy cycle (\ref{eq_14}) .

The enthalpy cycle is:
\begin{equation}
\left.
\begin{aligned}
\frac{d e_G}{dt} & \; = \;  - \: C_{(G,K)} \\
\frac{d e_K}{dt} & \; = \;  + \: C_{(G,K)}      \; + \: C_{(h,K)}       \;  - \:  D_K  \; \;  \\
\frac{d h}{dt}    & \; =   \hspace{19mm}    -   \: C_{(h,K)}    
                                        \: + \:  \left(
                                 \frac{1}{\rho} \;  \frac{\partial p}{\partial t} \: + \: Q 
                                                    \right) \; \; \\
\end{aligned}
\right\}
\label{eq_17}
\end{equation}

The first term of $G_h$ in (\ref{eq_16}), namely ${\rho}^{-1} \: \partial p /\partial t$, may be interpreted as the adiabatic expansion contribution to the cycle (in the absence of $\vec{\nabla}(p)$  or with $\vec{v} = 0$). 
The second term of $G_h$ in (\ref{eq_16}), namely $(1 - T_r/T) \: Q$, reduces to $Q$ in the enthalpy cycle(\ref{eq_17}). 
The efficiency factor $N_h = 1 - T_r/T$ which multiplies $Q$ in (\ref{eq_16}) is called the ``Carnot's factor related to a heat reservoir at temperature $T_r$'' in exergy theory (Borel 1987). 
$N_h$ is similar in form to the factor introduced by Pearce (1978), namely $1 - T_m(t)/T$.
 It is also one of the possible formulations for an efficiency factor considered by Lorenz (1967). 
However all the terms in (\ref{eq_14}) to (\ref{eq_16}) have a local physical meaning not given in the previous global studies where, for instance, only the integral of $N_h \: Q$ appeared. 
This formulation for $N_h$ was derived from the hypothesis that heating does not, in the long run, alter the average entropy (Lorenz 1967): this means that the space-time average of $Q/T$ is zero. 
Thus the space-time average of $Q$ is equal to the space-time average of $Q \: (1 - T_m/T)$ whatever $T_m$ is, provided it is a constant (cf. Pearce 1978).

Lorenz (1967) and Pearce (1978) have defined $T_m(r)$ so that, at time $t$, $1/T_m(t)$ is the average value of $1/T$ over the whole mass of the atmosphere. 
If so, the fundamental results are that the global-average efficiency factor vanishes and that the generation of available energy on a global average is zero for a uniform distribution of $Q$. 
Lorenz (1967) in his monograph interpreted spatial distributions of the efficiency factor $N_{\theta} = 1 - p^{\kappa}_{\theta}/p^{\kappa}$ as ``the effectiveness of heating at any point in producing APE'' ($p_{\theta}$ is
the average pressure at the isentrope passing through the point). 
Lorenz also showed a cross-section of $N_{\theta}$ (Lorenz 1967, Fig. 53). 
He introduced another efficiency factor when he defined the moist available energy (Lorenz 1978, 1979), interpreting it locally.

In the present study $T_r$ will be defined so that the local quantity appearing in $G_h$, namely $N_h$, is analogous to Pearce's efficiency factor. 
If so, heating (cooling) will produce available enthalpy where $N_h$ is positive (negative), which means $T> T_r$ $(T< T_r)$. 
Since $T_r$  must be a constant, let us define $1/T_r$ as the space-time average of the inverse of the temperature over a period $\Delta t = t_2 - t_1$ and over the whole atmosphere $\cal M$ having a total mass $M$:
\begin{equation}
\frac{1}{T_r} \: = \: 
\frac{1}{\Delta t} \int^{t_2}_{t_1} 
\left[\;
\iiint_{\cal M} \frac{1}{T} \; \frac{\mathrm{d}  m}{M} 
\;\right]
 \mathrm{d}  t
\; = \;
\frac{1}{\Delta t} \int^{t_2}_{t_1} 
\frac{1}{T_m(t)} \; \mathrm{d}  t
\: .
\label{eq_18}
\end{equation}
From (\ref{eq_18}), $1/T_r$ is also the time average over $\Delta t$ of the inverse of Pearce's temperature, namely of $1/T_m(t)$. The temperature $T_r$  is a characteristic value of the Earth's atmosphere.
It remains close to $250$~K in the present climatic conditions.

The fundamental properties resulting from the choice of $T_r$ are that the long term average generation of total available enthalpy vanishes for a uniform space-time average distribution of $Q$, and that the total entropy remains unchanged in the long run. 
Pearce (1978, Fig. 3) depicted the distribution of $N_h$: the general pattern of $N_h(\lambda, \varphi, p)$ is the same as that of $T(\lambda, \varphi, p)$.

      \subsection{The local energy law.} 
        \label{section_local_energy_law}

The total derivative of the quantity $e_G + e_K + a_h$ is easily obtained by adding the four equations of (\ref{eq_14}):
\begin{equation}
\frac{d}{dt}\left(  e_G + e_K + a_h \right)
\; = \;  
G_h - D_K
\; = \;
\frac{1}{\rho} \; \frac{\partial p}{\partial t}
\; + \left(  1 \: - \: \frac{T_r}{T}\right)
T \: \frac{ds}{dt} 
\; + \: 
\vec{v} \: . \:\vec{F} 
\: .
\label{eq_19}
\end{equation}
This quantity $e_G + e_K + _ah$ has been called specific total over-coenthalpy by Borel (1987).
The total derivative (\ref{eq_19}) vanishes for a frictionless and isentropic steady flow, for which
$\vec{F} = \vec{0}$, $Ds/Dt = 0$ and $\partial p/\partial t = 0$. 
Therefore, for such a flow, 
\begin{equation}
e_G + e_K + a_h = {\cal K} =  \mbox{constant}
\label{eq_20}
\end{equation}
along any particular streamline, though the constant ${\cal K}$ may vary from one streamline to
the next. 
This corresponds to the Bernoulli theorem for $e_G + e_K + a_h$.

Since from (\ref{eq_1}) and (\ref{eq_3}) the specific available enthalpy can be written as $a_h = h - a_N(\theta)$  where $a_N(\theta) = h_r + c_p \: T_r \: \ln(\theta/\theta_r)$,  Eq.(20) can be rewritten as $e_G + e_K + h = {\cal K} + a_N(\theta)$ where $a_N(\theta) = $constant.
Thus the choice of $a_h$ simply modifies the constant in the usual Bernoulli's law related to the total enthalpy $e_G + e_K + h$.

Rant has already defined the concept of  ``anergy'' (see Haywood 1974) as $h_r + T_r \:  (s - s_r) = a_N(\theta)$. 
Therefore, $a_N(\theta)$  is the specific anergy of the parcel, and, as expected by Rant, it is an untransformable part of energy (under Bernoulli's law in the present study). 
This concept of anergy has not been widely applied in the exergy literature. 
This peculiar physical application for $a_N(\theta)$ together with the local energy cycle (\ref{eq_14}) show that the mathematical concept of available enthalpy leads to coherent hydrodynamical results.

The Bernoulli theorem (\ref{eq_19}) can be obtained by subtracting $T_r \: ds/dt$ from the usual Bernoulli law derived from the enthalpy cycle (\ref{eq_17}). 
Then the usual Bernoulli law becomes $e_G + e_K + (h - T_r \: s) =$ constant, provided $T_r$ is a constant. 
This method can be considered as an alternative starting point for the introduction of the quantity $a_h$ defined by (\ref{eq_3}).\footnote{\color{blue} 
The comments of Dutton (1992) are discussed in Appendix~C.}
Nevertheless, since exergy is a general thermodynamic quantity (interest in which has grown in many parts of physics in recent years), the function $a_h$ has been chosen as a more rational starting point in the present paper.

      \subsection{The global conservation law.} 
        \label{section_global_conservation_law}

The integrals of various energies taken over the whole atmosphere are denoted here by capital letters. 
The kinetic energy of the horizontal wind is denoted by $E_K^{\star}$. 
Lorenz (1955) has shown that the total available potential energy (APE) obeys the following global conservation law (the total derivative of a global integral, a function of $t$ only,
reduces to a natural derivative with respect to time denoted by $d/dt$):
\begin{equation}
\frac{d}{dt}
\left( E_K^{\star} + \mbox{APE} \right) \: = \:  0 \: .
\label{eq_21}
\end{equation}
This property only holds for a hydrostatic atmosphere ($\partial \phi / \partial p = - 1/\rho$) with isentropic motions and without friction or topography (this latter assumption arises from the definition of APE).

In order to derive the global conservation law obeyed by $A_h$, it is convenient to integrate the local equation (\ref{eq_19}) over the whole mass of the atmosphere. 
Use of the equation of state and the continuity equation:
\begin{equation}
   \frac{1}{\rho} \;  \frac{\partial p}{\partial t} 
 \; + \: 
\overrightarrow{\nabla} . \: \vec{v}
 \; = \:  0 \: 
\label{eq_22}
\end{equation}
leads to
\begin{equation}
\frac{1}{\rho} \;  \frac{\partial p}{\partial t} 
 \; = \;
 \frac{d}{d t} 
\left(  R \: T\right)
 \; - \: 
\frac{1}{\rho} \;
\overrightarrow{\nabla} . \left(  p \: \vec{v} \right) \: .
\label{eq_23}
\end{equation}
Since the integral over the volume of the whole atmosphere of a divergence term vanishes (assuming that the velocity component normal to the uneven earth's surface is zero) the integral of (\ref{eq_23}) over the mass of the atmosphere (the integral element $dm = p d\tau$) is simply the time derivative of the integral of $R \: T$ over the mass, that is, the time derivative of $H - E_i$ (because $R = c_p - c_v \Rightarrow R\:T= H - E_i$). 
Therefore, for a frictionless isentropic flow and without the steady-flow assumption, the global integral of (\ref{eq_19}) gives
\begin{equation}
\frac{d}{dt}
\left( E_K + A_h \right) 
\: + \: 
\frac{d}{dt}
\left[ \: E_G - ( H - E_i ) \: \right]
\; = \:  0 \: .
\label{eq_24}
\end{equation}

For a hydrostatic atmosphere without topography ($z = 0$ at the surface), it is well known that $H - E_i$ is equal to $E_G$ (Lorenz 1967). 
But even if the atmosphere is nearly in hydrostatic equilibrium, the second time-derivative in (\ref{eq_24}) does not vanish identically when orography exists, although it must be small. 
The real hydrostatic global conservation law will be derived later in section~\ref{section_hydro_local_cycle}, starting from a hydrostatic local Bernoulli's law.

The exact global conservation law (\ref{eq_24}) explains the adjective ``available'' given to $A_h$: under the previous assumptions, an increase (decrease) in $A_h$ corresponds to a decrease (increase) in $E_K + (E_G + E_i - H) = E_K$. 
This can be compared with the definition of the exergy given by Baehr or Rant (see Haywood 1974): exergy is that part of (thermodynamic) energy that can be transformed into any other form of energy. 
Like the APE, the global available enthalpy can be understood from (\ref{eq_24}) to be a measure of the amount of energy available for conversion in $E_K + (E_G + E_i - H)$ under isentropic flow (the last quantity in parentheses vanishing under the assumption of hydrostatic equilibrium and without orography).

This global conservation law (\ref{eq_24}) is analogous to the first of the four necessary conditions given by Lorenz to define a meteorological available energy. 
These conditions are (Lorenz 1955):
\begin{itemize}[label=,leftmargin=3mm,parsep=0.1cm,itemsep=0cm,topsep=0cm,rightmargin=2mm]
\vspace*{-1mm}
\item 1) APE$ \: + \: E_K^{\star}=$ constant under adiabatic flow.
\item 2) APE is completely determined by the distribution of mass.
\item 3) APE$\:=0$ if the stratification is horizontal and statically stable.
\item 4) APE$\: >0$ if the stratification is not both horizontal and statically stable.
\end{itemize}
Similarly $A_h$ is completely determined by the distribution of mass, and does not depend upon the momentum distribution-condition 2 of Lorenz. 
The third condition is the choice of the ``reference state'' for which $A_h$ cancels out. 
If the global integral of $a_p$ is zero, the reference state is the one for which $A_T = 0$. 
From (\ref{eq_5}), since ${\cal F}(X) = 0$ only for $X= 0 \:\; (\Rightarrow  T = T_r)$, $A_T$ vanishes only for the isothermal atmosphere at temperature $T_r$.
This is the reference state introduced by Pearce (1978) and Dutton (1973, 1976). 
Assuming that this condition, $A_p = 0$, is valid on a long-term average, the fourth condition of Lorenz becomes with the available enthalpy formulation an approximate one, only valid on a long-term average: $A_p \approx  0 \Rightarrow A_h = A_T > 0$ if $T \neq T_r$, anywhere. 

We thus obtain the following definition for $p_r$: $\ln(p_r)$ is the space-time average of $\ln(p)$, whatever the topography is. 
It is a definition similar to the one for $T_r$ (\ref{eq_18}):
\begin{equation}
\ln({p_r}) \: = \: 
\frac{1}{\Delta t} \int^{t_2}_{t_1}
\left[\;
\iiint_{\cal M} \ln(p) \; \frac{\mathrm{d}  m}{M} 
\;\right]
 \mathrm{d}  t
\: .
\label{eq_25}
\end{equation}
It is worth noting that for the earth's atmosphere $p_r \approx \:$constant for practical purposes, whereas for a limited time interval $T_r$ may not be (owing to the diurnal cycle). Typically $p_r \approx  p_{00}/e = 367.88$~hPa, where $p_{00}/e$ is the pressure for the scale height of any isothermal and hydrostatic atmosphere, provided that the surface pressure is $p_{00}$, (Gill 1982, p. 49).

The available enthalpy concept can thus be seen as a generalization of the available energy of Pearce rather than that of Lorenz. 
As anticipated by Pearce, the term ``reference state'' is not relevant, since even if this isothermal atmosphere corresponds to zero $A_h$ in
a long-term average, it cannot be reached globally from the actual atmosphere through any isentropic redistribution: all the available enthalpy reservoir cannot be converted into kinetic energy (that would be unrealistic). 
Equation (\ref{eq_24}) is only a global conservation law; the local capacity to general $e_K$ is explained rather by the local cycle (\ref{eq_14}). 
This again is in agreement with Pearce's (1978) concluding remarks.

As in thermodynamic exergy theory, Dutton identified the associated isothermal reference state with the natural equilibrium state of the atmosphere, and showed it to
be the stable state in Gibbs' sense (it has the maximum entropy of all states with same total mass $M$ and total energy $H$). 
The purpose of this paper however is not to study such properties of the reference state, which anyhow may not be found with the available enthalpy formulation (as will be explained in section~\ref{section_summary}).
We will be concerned rather with the local conversions of energy into its various forms.\footnote{\color{blue} 
In fact, it is meaningless to define a reference state as being really ``attainable'' from the actual one, for instance by adiabatic process.
The method used by Margules or Lorenz has the fundamental drawback to be sensible to the appearance of a parcel of air with a smallest value of entropy or $\theta$ close to the poles.
In this case, the stratified reference state is immediately modified everywhere on earth, included in the Tropics!
This unrealistic process may be called the ``cold penguin problem''...}

From this section it follows that the available enthalpy is a state function which satisfies not only the four aforementioned conditions, but also the condition that the sum of the specific mechanical energy ($e_G + e_K$) and the specific available enthalpy ($a_h$) be subject to Bernoulli's conservation law (\ref{eq_20}).
The ``specific available'' energies that could be associated with the global available energies of Lorenz or Pearce do not satisfy this last condition.

      \subsection{Potential change in total entropy.} 
        \label{section_potential_entropy}

It is usually accepted (Keenan 1951) that, in order to apply the thermodynamic concept of exergy to a system, the system must be surrounded by a heat reservoir at constant temperature and pressure (the so-called thermostat), not definable in meteorology\footnote{\color{blue} 
See the footnote ${}^{\mbox{\tiny\ref{footnote_Keenan}}}$}.
However, considering the results of section~\ref{section_physical_properties}, it appears that the concept of available enthalpy is relevant to atmospheric energetics.


The algebraic structure (\ref{eq_3}) for $a_h$ can be introduced starting from a more meteorological approach.\footnote{\label{footnote_Gibbs}\color{blue} 
In fact, this corresponds to the approach of Gibbs (1873b).}
It is demonstrated in this section that $a_h$ can indeed be related to what can be called a {\it potential change in total entropy \/}. 
This new approach is similar to the link between the entropy state function and the potential temperature concept in meteorology.

It is possible first to derive from general thermodynamics the fact that $a_T$ is a positive quantity.
 Let us consider a thermodynamic system which consists of a heat reservoir at constant temperature $T_r$ (the thermostat) and a unit-mass particle of an ideal gas (the parcel). 
Suppose this system undergoes the following process. 
The temperature of the parcel changes from $T$ to $T_r$ through an irreversible and isobaric heat transfer with the thermostat. 
It is easy to demonstrate that the change in total entropy of the system is exactly $a_T/T_r$.

As usual in such thermodynamic computations the property that entropy is a state function (an exact differential) may be used to imagine an associated reversible path which links the end states of the parcel: $(T, p)$ and $(T_r, p)$. 
For the parcel, let the associated reversible process be that isobaric change from $T$ to $T_r$. 
For an elementary step in such a process, $ds = c_p \: dT/T$ and the change in entropy between the end states of
the parcel is $- \: c_p \: \ln(T/T_r)$. 
During the irreversible process the thermostat has supplied to the parcel the heat quantity $q = + \: c_p \: (T_r - T)$, and its temperature remains constant. 
Thus the change in entropy of the thermostat is simply $- q / T_r = c_p \: (T- T_r)/T_r$ (i.e. the heat received divided by the constant temperature). 
Adding these two changes in entropy yields  $a_T/T_r = c_p \: {\cal F}(T/T_r-1)$.

Therefore, since  $a_T/T_r$ is the change in total entropy due to the irreversible process described above,  $a_T$ is positive by virtue of the thermodynamic principle of increase of entropy: in any natural process taking place within an isolated system, the total entropy either increases or remains constant.
It is also possible to demonstrate that $a_T$ is positive from (\ref{eq_6}) and by using pure mathematics, by virtue of ${\cal F}(X) \geq 0$.

Likewise, since from (\ref{eq_4}) and (\ref{eq_7}) ah/T, can be put in the form
\begin{equation}
\frac{a_h}{T_r}  \: = \: \frac{a_T}{T_r} + R \:\ln\left( \frac{p}{p_r} \right) 
\:
\label{eq_26}
\end{equation}
and if the thermodynamic state of the parcel is defined by its temperature and its pressure, the quantity $a_h/T_r$ is exactly the change in total entropy of the previous system (parcel + thermostat) when the parcel undergoes the following processes:
\begin{itemize}[label=,leftmargin=3mm,parsep=0.1cm,itemsep=0cm,topsep=0cm,rightmargin=2mm]
\vspace*{-1mm}
\item i) a change from $(T, p)$ to $(T_r, p)$ through an irreversible and isobaric heat transfer with the thermostat, yielding $a_T/T_r$.
\item ii) a change from $(T_r, p)$ to $(T_r, p_r)$ through a sudden and irreversible adiabatic and isothermal process, yielding $a_p/T_r = R \: \ln(p/p_r)$.
\end{itemize}
This latter computation is made, as usual, considering once more an associated reversible and isothermal path between the end states $(T_r, p)$ and  $(T_r, p_r)$. 
The change in entropy of the parcel is thus $+R \: \ln(p/p_r) = a_p/T_r$. 
And since the thermostat does not exchange heat with the parcel, its change in entropy is zero during the irreversible process (ii).

Let us define
\begin{equation}
\Delta S^0(T,p)  \: = \: \frac{a_h}{T_r}
\: 
\label{eq_27}
\end{equation}
as the potential change in total entropy which only depends on the two state functions $T$ and $p$. 
This introduction of $\Delta S^0$  is similar to the definition of the potential temperature. 
The definition of $\Delta S^0$ needs the two processes (i) and (ii) associated with the two
numerical constants $T_r$ and $p_r$. 
The definition of $\theta$ only needs one process and one numerical constant $p_{00}$.

The specific available enthalpy can thus be expressed in terms of $\Delta S^0$  as
\begin{equation}
a_h (T, p)  \: = \: T_r \:  \Delta S^0
\: .
\label{eq_28}
\end{equation}
When $a_h$ is defined by (\ref{eq_28}), there is no need to seek any real outer medium of constant temperature as must be done in exergy theory. 
The thermostat involved in the process (i) is a mere imaginary one. 
Landau and Lifchitz (1958, section~20) have derived an expression analogous to (\ref{eq_27}), but they gave to this thermostat the usual physical interpretation (the temperature of the outer medium). 
Feidt (1987) suggested defining $T_0$  in the extrinsic function $H - T_0 \:  S$ as a space-time average of the outer temperature.
In Eq.~(\ref{eq_28}), $T_r$ is defined rather as the space-time average (\ref{eq_18}) of the fluid temperature.
Thus (\ref{eq_18}) and (\ref{eq_28}) define $a_h$ as an intrinsic state function and justify its local physical introduction and interpretation in atmospheric energetics.

 \section{HYDROSTATIC ENERGETICS ON AN ISOBARIC AVERAGE.} 
\label{section_hydro_energetics}

The available enthalpy theory must be expressed under the hydrostatic assumption in order to study further connections with the theories of Lorenz and Pearce and to show the possible applications to numerical modelling or data analysis. 
The kinetic energy $e_K = \vec{v} \: . \:  \vec{v}  \:/ \:2$  must be replaced by the corresponding hydrostatic kinetic energy of the horizontal wind denoted by an asterisk $e_K^{\ast} = \vec{u} \: . \:  \vec{u}  \:/\:2$.
Indeed, only $e_K^{\ast}$ and not $e_K$  leads to the hydrostatic conservation laws demonstrated later on.
More generally, all terms which are modified under the hydrostatic assumption will be denoted by an asterisk.

      \subsection{Hydrostatic local cycle.} 
        \label{section_hydro_local_cycle}

Using the total derivatives (\ref{eq_10}), (\ref{eq_11}) and (\ref{eq_13}) and introducing the continuity equation $\overrightarrow{\nabla}_{\!p} \: . \: u + \partial \omega /  \partial p = 0$ in the total derivative of $e_K^{\ast}$  (the horizontal vector operator $\overrightarrow{\nabla}_{\!p}$, refers to a differentiation at constant pressure, see Kasahara 1974), the hydrostatic local cycle is
\begin{equation}
\left.
\begin{aligned}
\frac{d e_G}{dt} & \; = \;  - \: C_{(G,K)} \\
\frac{d e^{\ast}_K}{dt} & \; = \;  + \: C_{(G,K)}      \;  + \: C^{\ast}_{(h,K)}    \: \;  + \:  G^{\ast}_K   \;  - \:  D^{\ast}_K  \; \;  \\
\frac{d a_T}{dt} & \; = \;  + \: C_{(p,T)}   \: \;  -   \: C^{\ast}_{(h,K)}   \: \;  + \:  G^{\ast}_h \; \; \\
\frac{d a_p}{dt} & \; = \;  - \: C_{(p,T)} 
\end{aligned}
\right\}
\label{eq_29}
\end{equation}
The conversion terms $C_{(G.K)}$ and $C_{(p.T)}$ are the same as in the local cycle (\ref{eq_14}), whereas the hydrostatic terms are the usual ones
\begin{equation}
\left.
\begin{aligned}
C^{\ast}_{(h,K)} & =   - \: \frac{R}{p} \; T \; \omega 
\; \; \; \; \; \; \; \;
D^{\ast}_K \: = \: - \:  \vec{u} \: .  \: \vec{F} 
 \\
G^{\ast}_K  & =     + \: \frac{1}{\rho} \: \frac{\partial p}{\partial t}
\; \; \; \; \; \; \; \; \; \; \;
 G^{\ast}_h \: = \: 
    \left(  1 - \frac{T_r}{T} \right) \: Q
\end{aligned}
\right.
\label{eq_30}
\end{equation}
The hydrostatic cycle can also be written in terms of $e_K^{\ast}$ and $a_h$ alone:
\begin{equation}
\left.
\begin{aligned}
\frac{d e^{\ast}_K}{dt} & \; = \;  - \: B(\phi)      \;  + \: C^{\ast}_{(h,K)}    \: \;  - \:  D^{\ast}_K  \; \;  \\
\frac{d a_h}{dt}             & \; = \;\;\;\;\;\;\;\;\;\;\;\;\;\;\:  -   \: C^{\ast}_{(h,K)}  \: \: \;  + \:  G^{\ast}_h \; \; \\
\end{aligned}
\right\}
\label{eq_31}
\end{equation}
where
\begin{equation}
B(\eta) \;  = \;  \overrightarrow{\nabla}_{\!p} \: . \:  ( \eta \: \vec{u} \, ) + \frac{\partial }{\partial  p}( \eta \: \omega  )
\label{eq_32}
\end{equation}
is the isobaric flux divergence term for any scalar $\eta$ ($\eta = \phi$ in (\ref{eq_31})). 
The choice of the subscripts of $C^{\ast}_{(h.K)}$ and $G^{\ast}_{h}$ is made according to this cycle (\ref{eq_31}); it is kept throughout the study even when $a_h$ is split into $a_T$ and $a_p$ as in (\ref{eq_29}). 
If $\eta$ is a scalar quantity, the integral of $B(\eta)$ over the mass of an open domain extending vertically between two isobaric surfaces is simply related to the flux of $\rho \: \eta \: \vec{u}$ through the lateral vertical boundaries and to the integral of $\eta \: \omega /g$ over the two isobaric surfaces.

As in sections~\ref{section_local_energy_law} and \ref{section_global_conservation_law}, it is easy to prove that, under the hydrostatic assumption, $e_G + e^{\ast}_K + a_h$ is subject to a Bernoulli conservation law, whereas the sum $E^{\ast}_K + A_h$ is exactly conserved for an isentropic flow above a flat earth, as shown by Lorenz (1955).

      \subsection{Energy components.} 
        \label{section_energy_components}

Let us consider an open atmospheric domain. 
The isobaric average of any scalar quantity $\eta$ is defined as
\begin{equation}
\overline{\eta}(p) \;  = \;  \iint_{\cal S} \eta(\lambda, \varphi, p) \; \frac{d\Sigma}{{\cal S}}
\label{eq_33}
\end{equation}
where $d\Sigma = r^2 \cos(\varphi) \: d\lambda \: d\varphi$ is the element of horizontal area. 
The integral extends over an isobaric surfaced located within the region, the area of its horizontal projection being $\cal S$. 
Actually, $d\Sigma$  is the horizontal projection of the element of the isobaric surface, not the real surface area of this element.

Introducing the isobaric average of the temperature, $\overline{T}$, the temperature energy component $a_T$ can be divided into the sum of the three energy components $a_B$, $a_S$ and $a_C$:
\begin{equation}
\left.
\begin{aligned}
a_T  & \; = \;  a_B \; + \; a_S \; + \; a_C \; , \\
a_B  & \; = \;  c_p \: T_r \: {\cal F}(X_B) \; , \;\;\;\;\; X_B \; = \; \frac{(T - \overline{T})}{\overline{T}} \; , \\
a_S  & \; = \;  c_p \: T_r \: {\cal F}(X_S) \; , \;\;\;\;\; X_S \; = \; \frac{(\overline{T} - T_r)}{T_r} \; , \\
a_C  & \; = \;  c_p \: T_r \: X_S \; X_B \: .
\end{aligned}
\right.
\label{eq_34}
\end{equation}

The component $a_B$, which depends on variations of the temperature on an isobaric surface, is called the baroclinicity energy component of $a_h$ (Pearce 1978). 
The component $a_S$ depends on variations in the vertical profile of $\overline{T}$, and is called the static stability component of $a_h$ (for an isothermal atmosphere at $T_r$, $a_S$ vanishes). 
It is important to notice that, although the complementary component $a_C$ is zero on an isobaric average ($\overline{X_B} = 0$ and $X_S = $constant), it does not vanish locally.

It must be pointed out that the three specific energy components $a_B$, $a_S$ and $a_C$ are not intrinsically defined since they require the definition of $\overline{T}$: they all depend on the location as well as the spatial extent of the limited-area atmospheric domain. 
Only $a_T$ and $a_p$ possess an intrinsic and local definition. 
Nevertheless, the isobaric averages $\overline{a_B}$ and $\overline{a_S} = a_S$ are really related to a particular (limited-area) isobaric layer.

After a Taylor series expansion of ${\cal F}(X)$  about $X= 0$, only the leading term, i.e. $X^2/2$, need be retained because typical values of $|X|$, $|X_B|$ and $|X_S|$  are not greater than $0.25$ in the atmosphere. 
Therefore, replacing ${\cal F}(X)$ by this leading order approximation, the energy components $a_T$, $a_B$ and $a_S$ reduce with good accuracy to the energy components of Pearce (1978) denoted by $\widetilde{a}$, $\widetilde{a}_B$ and $\widetilde{a}_S$:
\begin{equation}
\left.
\begin{aligned}
a_T  & \; \approx \;  c_p \: \frac{(T - T_r)^2}{2 \: T_r} \; \equiv \; \widetilde{a} \; , \\
a_B  & \; \approx \; \left(\frac{T_r}{\overline{T}}\right)
            \: \left[ \;
            c_p \: \frac{(T - \overline{T})^2}{2 \: \overline{T}} 
            \; \right] 
            \; \equiv \; \left(\frac{T_r}{\overline{T}}\right)^2 \: \widetilde{a}_B 
\\
\mbox{where}\; \; 
\widetilde{a}_B & \equiv \; c_p \: \frac{(T - \overline{T})^2}{2 \: T_r}  
\; , \\
a_S  & \; \approx \;  c_p  \: \frac{(\overline{T}-T_r)^2}{2 \: T_r}  
            \; \equiv \; \widetilde{a}_S \; . 
\end{aligned}
\right.
\label{eq_35}
\end{equation}
Since the value of term $(T_r/\overline{T})^2$ is close to unity, $a_B$ can be indeed approximated by $\widetilde{a}_B$ in (\ref{eq_35}).

The available enthalpy components $a_B$, $a_S$, $a_C$ and $a_p$ provide a generalization of Pearce's available energy components $\widetilde{a}_B$ and $\widetilde{a}_S$  (global averages of $a_C$ and $a_p$ vanish).
Nevertheless, it must still be demonstrated that these four available enthalpy components lead to a new coherent energy cycle which generalizes that of Pearce. 
This is done in the next section.

      \subsection{Hydrostatic cycle of isobaric average components.} 
        \label{section_hydro_cycle_components}

According to (\ref{eq_29}) or (\ref{eq_31}), the conversion term $C^{\ast}_{(h,K)}$ allows the available enthalpy to supply frictional energy dissipation in the long-term average (Lorenz 1967), thus explaining the maintenance of the general circulation (i.e. the kinetic energy reservoir).
If $\eta'$  denotes a deviation from the isobaric average for any scalar quantity $\eta$, the isobaric average of $C^{\ast}_{(h,K)}$ is
\begin{equation}
\overline{C^{\ast}_{(h,K)}}  
 \; = \; 
- \: \frac{R}{p} \;\; \overline{\, \omega \: T\,} 
 \; = \; 
- \: \frac{R}{p} \; 
  \left(\;
     \overline{\omega} \; \overline{T} 
      \: + \:
     \overline{ \omega' \: T'} 
 \; \right)
\label{eq_36}
\end{equation}

The problem, as pointed out by Saltzman and Fleisher (1960), is that $\overline{\omega}$ does not vanish over a limited space region: $\overline{\omega}$ is zero only on an entire and closed isobaric surface. 
The mass of the hydrostatic atmosphere which is located above this entire isobaric surface is indeed equal to $p\: {\cal S} / g$. 
It is a constant and its total derivative vanishes: it is the integral of $\omega$ over the isobaric surface.
Moreover, typical values show that $\overline{\omega} \: \overline{T}$ and $\overline{\omega' \: T'}$ are at least of the same order of magnitude for a synoptic-scale region (because $\overline{\omega} \sim 0.1 \: | \, \omega' \, | $ but $\overline{T} \sim 10 \: | \, T ' \, | $).

The global theories of Lorenz and Pearce made use of the property $\overline{\omega} = 0$, and the term $- R \: \overline{\omega} \: \overline{T} / p$ could not appear in their available energy equations, although the opposite term $+ R \: \overline{\omega} \: \overline{T} / p$ appeared in the kinetic energy equation. 
It is thus impossible to redress this imbalance if the approach of Lorenz or Pearce is chosen.
It is an important advantage of the available enthalpy approach to be able to manage all terms depending on $\overline{\omega}$.

Nevertheless, several studies on limited-area energetics have used the formulae of Lorenz and have dropped the term $+ R \: \overline{\omega} \: \overline{T} / p$ in the kinetic energy equation in order to ensure a balance with the ``truncated and local'' APE equation (see e.g. Muench 1965; Brennan and Vincent 1980; Michaelides 1987). 
Taking account of $\overline{\omega} \neq 0$ in the conversion term $\overline{C^{\ast}_{(h,K)}}$  is the challenge for the definition of energetics of isobaric average components.
This is possible with the available enthalpy $a_h$, whereas it was not possible with the theories of Lorenz or Pearce.

The hydrostatic cycle of isobaric average energy components is obtained as a set of isobaric local derivatives with respect to time of $\overline{e_K^{\ast}}$,  $\overline{a_B}$ and $\overline{a_S}$. 
Each budget equation has the generic form:
\\
\hspace*{2cm}Isobaric rate of change = flux convergences + conversions + sources/sinks.

Under the hydrostatic assumption, the use of the isobaric rates of change (still denoted by $\partial/\partial t$ in this section) together with the isobaric flux divergences of $e_K^{\ast}$,  $a_B$,  $a_S$,  $a_C$ and  $a_p$ (isobaric flux divergences $=B(\eta)$ terms) leads to the following energy equations of the isobaric average: 
\begin{equation}
\left.
\begin{aligned}
\frac{\partial \,\overline{e^{\ast}_K}}{\partial t} & 
 \; = \; - \: \overline{B(e^{\ast}_K)} 
      \; - \: \overline{B(e_G)}
      \; + \: \overline{C_{(B,K)}} 
      \; + \: \overline{C_{(S,K)}}  \hspace*{17mm}
      \; - \:  \overline{D^{\ast}_K} \\
\frac{\partial \,\overline{a_B}}{\partial t} & 
 \; = \; - \: \overline{B(a_B)}     \hspace*{17mm}
      \; - \: \overline{C_{(B,K)}}  \hspace*{17mm}
      \; + \: \overline{C_{(S,B)}} 
      \; + \: \overline{G_B} \\
\frac{\partial\, \overline{a_s}}{\partial t} & 
 \; = \; - \: \overline{B(a_S)}  
      \; - \: \overline{B(a_C)} 
      \; - \: \overline{B(a_p)}     \hspace*{1mm}
      \; - \: \overline{C_{(S,K)}}
      \; - \: \overline{C_{(S,B)}}
      \; + \: \overline{G_S}
\end{aligned} \; \;
\right\}
\label{eq_37}
\end{equation}
where
\begin{equation}
\left.
\begin{aligned}
 \overline{C_{(B,K)}} & 
 \; = \; - \: \frac{R}{p} \; \overline{ \omega' \: T'} 
 \; , \hspace{5mm}
 \overline{C_{(S,K)}} 
 \; = \; - \: \frac{R}{p} \; \overline{\omega} \; \overline{T}  \,  , \\
 \overline{C_{(S,B)}} & 
 \; = \;
  \left[ \:
     \left(  1 - \frac{T_r}{\overline{T}} \right)
      \: + \: 
      \frac{c_p}{R} \: 
      \frac{T_r}{\overline{T}}
      \frac{\partial \, \ln(\overline{T})}{\partial \,\ln(p)}
   \; \right] \; \overline{C_{(B,K)}} \,  ,   \\
 \overline{G_B} & 
 \; = \; \frac{T_r}{\overline{T}}\;
          \overline{
              \left(  1 - \frac{\overline{T}}{T} \right) \: Q
                   }
 \; , \hspace{5mm}
 \overline{G_S} 
 \; = \; \left(  1 - \frac{T_r}{\overline{T}} \right) \; \overline{Q} \,  .
\end{aligned} \; \;
\right.
\label{eq_38}
\end{equation}
and where the isobaric averages of the isobaric flux divergences of $a_C$ and  $a_p$  can be written as
\begin{equation}
\left.
\begin{aligned}
 \overline{B(a_C)} & 
 \; = \;
 \; c_p \;
     \left(  1 - \frac{T_r}{\overline{T}} \right)
  \; \left( \:
      \overline{B(T)} \: - \: \overline{\omega} \;
      \frac{\partial \, \overline{T}}{\partial \,p}
     \; \right)
 \; - \;  
     \left[ \:
      \frac{c_p}{R} \: 
      \frac{T_r}{\overline{T}}
      \frac{\partial \, \ln(\overline{T})}{\partial \,\ln(p)}
      \; \right]  \; \overline{C_{(B,K)}} \,  ,  \\
 \overline{B(a_p)} & 
 \; = \;
 + \: \frac{R}{p} \; 
     \overline{\omega} \; T_r  \: .
\end{aligned} \; \;
\right.
\label{eq_39}
\end{equation}

The set of three equations (\ref{eq_37}) is similar to Pearce's (1978) cycle except for the new conversion term $\overline{ C_{(S,K)}}$  and the flux divergence terms $\overline{B(a_C)}$ and $\overline{B(a_p)}$  in the equation for $\overline{a_S}$. 
The term $\overline{ C_{(S,K)}}$ is that expected part of $\overline{C^{\ast}_{(h,K)}}$  which takes account of $\overline{\omega} \neq 0$. 
This exact cycle is appropriate for the study of the energetics of the isobaric layers of an open domain. 
The flux divergences ($B$ terms) of $e_G$, $e_K^{\ast}$,  $a_B$ and  $a_S$ appear because the region is not closed (open boundaries). 
Note that these flux divergence terms were implicit in Pearce's study (see his Eq. 15(a), p. 745).

Since $T_r/\overline{T} \approx 1$, the conversion term $\overline{C_{(S,B)}}$ is similar to the one introduced by Pearce, denoted by $\widetilde{C}_{(S,B)}$:
\begin{equation}
\left.
\begin{aligned}
 \overline{C_{(S,B)}} & 
 \; = \;
  \left[ \:
     \left( \frac{T_r}{\overline{T}} \right)\;
     \left( \frac{\overline{T}-T_r}{T_r} \right)
      \: + \: 
     \left( \frac{T_r}{\overline{T}} \right)^2\;
      \frac{c_p}{R} \: 
      \frac{p}{T_r}
      \frac{\partial \, (\overline{T}-T_r)}{\partial \, p}
   \; \right] \; \overline{C_{(B,K)}}  \,  ,  \\
 \widetilde{C}_{(S,B)} & 
 \; = \;
  \left[ \:
     \hspace{12mm}
     \left( \frac{\overline{T}-T_r}{T_r} \right)
      \: + \: 
     \hspace{14mm}
      \frac{c_p}{R} \: 
      \frac{p}{T_r}
      \frac{\partial \, (\overline{T}-T_r)}{\partial \, p}
   \; \right] \; \overline{C_{(B,K)}} 
 \: .
\end{aligned} \; \;
\right.
\label{eq_40}
\end{equation}

The generation terms $\overline{G_B}$, and $G_S$ can also be related to the associated generation terms of Pearce, denoted by $\widetilde{G}_B$ and $\widetilde{G}_S$. 
Using $1 - \overline{T}/T \approx (T - \overline{T})/\overline{T}$, and then introducing $\overline{Q}$, $\overline{G_B}$ can be approximated by
\begin{equation}
 \overline{G_B} 
 \; \approx \;
     \left( \frac{T_r}{\overline{T}} \right)^2\;
    \widetilde{G}_B 
    \; \; \; \;\mbox{where}\; \; 
    \widetilde{G}_B 
    \; \equiv \;
     \overline{
     \left( \frac{T - \overline{T}}{T_r} \right)
     \left( Q - \overline{Q} \right)
     }
 \: .
\label{eq_41}
\end{equation}
Likewise, introducing the global average of $Q$ (denoted by $\widehat{Q}$),  $\overline{G_S}$ can be related to the generation of Pearce $\widetilde{G}_S$ by
\begin{equation}
 \overline{G_S} 
 \; = \;
     \left( \frac{T_r}{\overline{T}} \right)\;
    \widetilde{G}_S 
    \: + \: 
     \left( 1 - \frac{T_r}{\overline{T}} \right)
    \; \widehat{Q}
    \; \; \; \;\mbox{where}\; \; 
    \widetilde{G}_S 
    \; \equiv \;
     \left( \frac{\overline{T} - T_r}{T_r} \right)
     \left( \overline{Q} - \widehat{Q} \right)
 \: .
\label{eq_42}
\end{equation}

The second term of $G_S$ which involves $\widehat{Q}$ roughly cancels out in the long term and global average (if changes in time in $\widehat{Q}$ are neglected) since $T_r$ is defined by (\ref{eq_18}). 
Thus, the global-average generation of $\overline{a_S}$ is nearly the same as in Pearce's study. 
But this is not true for a particular isobaric layer of an atmospheric domain of limited area where $\widehat{Q}$ must be retained in the second term of $G_S$. 
It is a new generation term.

      \subsection{Numerical evaluation of energy components.} 
        \label{section_numerical_energy_components}

\begin{figure}[t]
\centering
\includegraphics[width=0.6\linewidth,angle=0,clip=true]{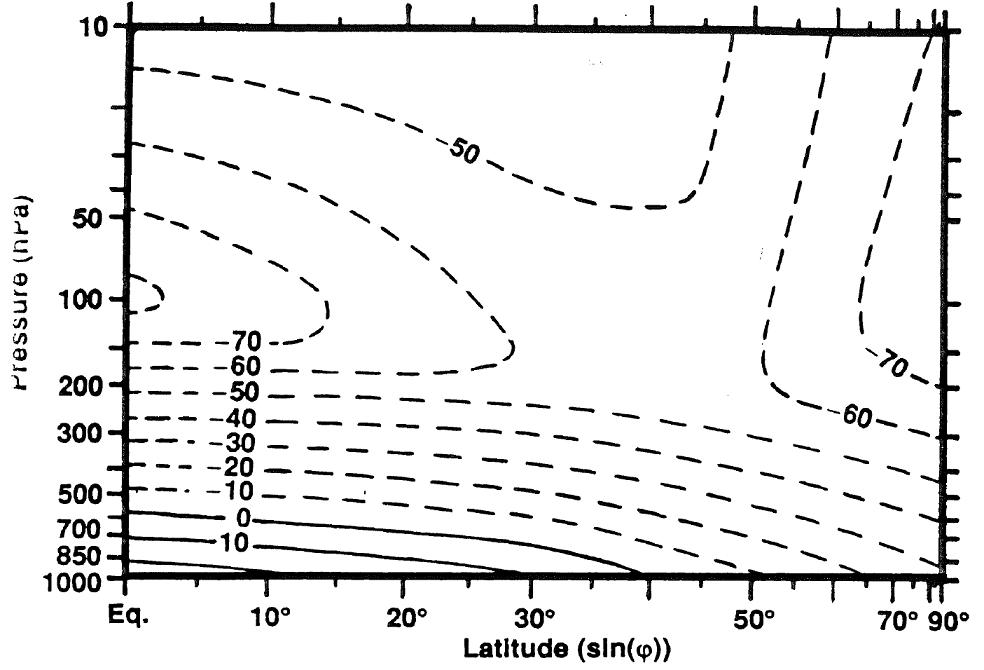}
\caption{
{\it
Cross-section of zonally averaged temperature in ${}^{\circ}$C corresponding to the northern hemisphere winter conditions ($\partial \, T / \partial \, z$ is continuous).
}
\label{Fig_1}}
\end{figure}

Using the hypothetical zonal-average temperature field depicted in Fig.\ref{Fig_1} (continuous and derivable algebraic vertical profiles), all the energy components $a_T$, $a_p$, $a_B$, $\widetilde{a}_B$, $a_S$, $\widetilde{a}_S$, $a_C$ and the component of Lorenz, $a_L$, have been computed taking $T_r = 250$~K. 
The ``specific'' Lorenz's approximate expression can be written as \footnote{\color{blue} Expressions (\ref{eq_43}) for $a_L$ and $\overline{\sigma}$ are written differently in Marquet (1991).
This formulation (\ref{eq_43}) for $a_L$ is more interesting because it allows easier comparison of $a_L$  with $a_B$ given by (\ref{eq_35}).
It is thus possible to recover the first-order approximation of the non-flow exergy availability function $a_B$, simply  by replacing $\overline{\sigma}$ in (\ref{eq_43}) by $(\overline{T}/T_r)^2$, differently from Lorenz's value $\overline{\sigma}(p)$.
It is also easier to understand that the adiabatic lapse rate ($\partial T/\partial z = - \: g/c_p$ or $\partial \theta /\partial z = 0$) leads to $\overline{\sigma} = 0$ and infinite value of $a_L$ in Lorenz's formulae, whereas $a_B$ remains finite with $\overline{\sigma}(p)$ replaced by $(\overline{T}/T_r)^2$.}
\begin{equation}
a_L 
\; = \;
    c_p \;
    \frac{\left( T - \overline{T} \right)^2}{2 \; \overline{T} \; \overline{\sigma}}
\; \; \; \;
\mbox{where}
\; \; 
    \overline{\sigma}(p)
    \; = \;
       1 
       \: - \:
       \frac{c_p}{R} \;
       \frac{\partial \ln(\overline{T})}{\partial \ln(p)}
    \; = \;
        - \:  \frac{c_p}{R} \;
       \frac{\partial \ln(\overline{\theta})}{\partial \ln(p)}
 \; .
\label{eq_43}
\end{equation}
Note that $a_L$ is not a specific quantity since the APE of Lorenz is not defined locally, even if the integral of $a_L$ over the whole mass is equal to APE. 
It is nevertheless interesting to compare the available enthalpy components with the ``reservoir'' $a_L$ which has been widely studied by itself (the term reservoir is restricted to positive energy components
such as $a_T$, $a_B$, $\widetilde{a}_B$, $a_S$, $\widetilde{a}_S$ or $a_L$).

\begin{figure}[t]
\centering
\includegraphics[width=0.6\linewidth,angle=0,clip=true]{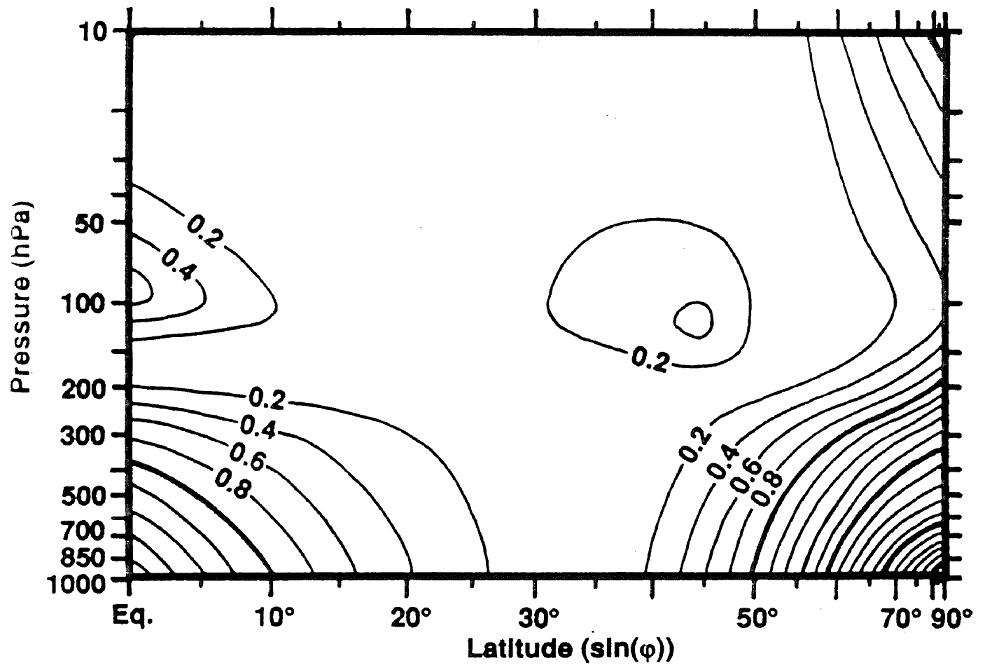}
\caption{
{\it
Distribution of $a_L$ (Lorenz specific available energy), in kJ~kg${}^{-1}$, corresponding to the temperature field depicted in Fig.1.
}
\label{Fig_2}}
\end{figure}

\begin{figure}[t]
\centering
\includegraphics[width=0.6\linewidth,angle=0,clip=true]{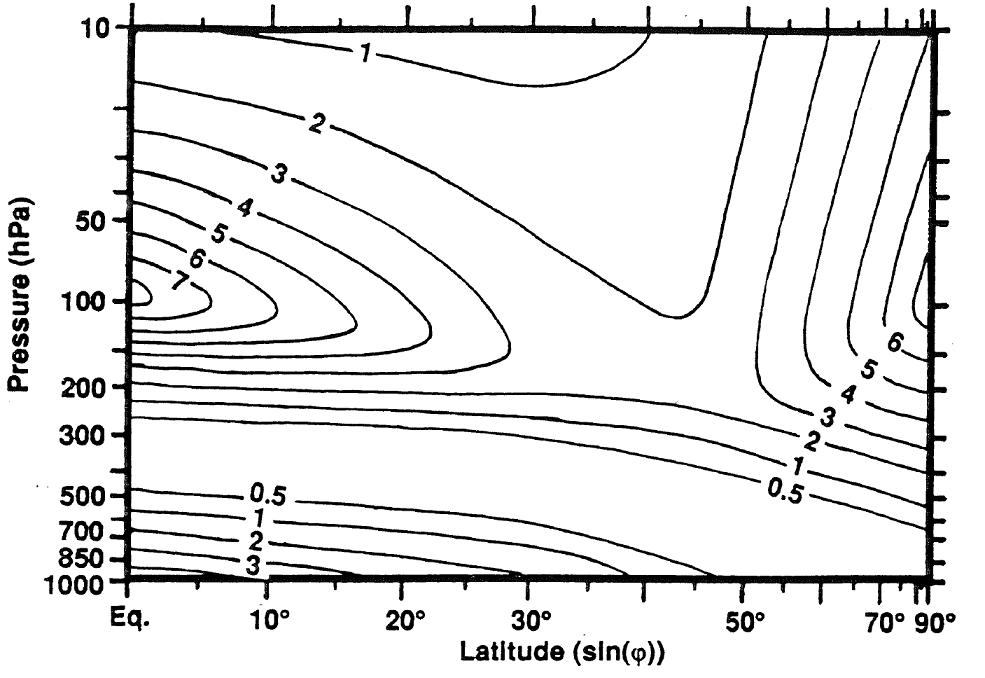}
\includegraphics[width=0.3\linewidth,angle=0,clip=true]{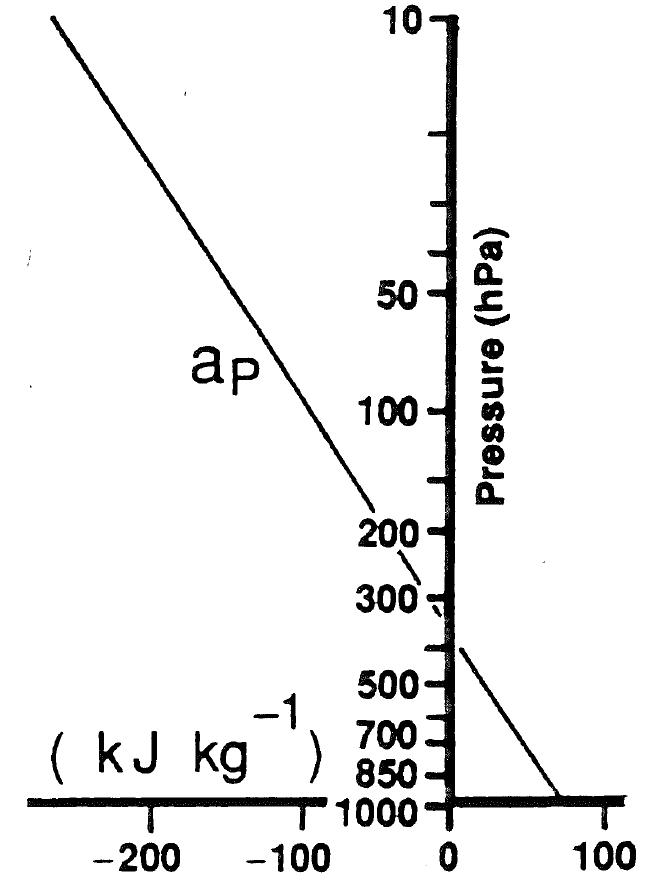}
\\ \hspace*{3cm} (a) \hspace*{7cm} (b)
\caption{
{\it
(a) Spatial distribution of $a_T$. (b) Vertical profile of $a_p$. Units are kJ~kg${}^{-1}$.
}
\label{Fig_3}}
\end{figure}

From Fig.\ref{Fig_2}, a typical value of $a_L$ is $0.5$~kJ~kg${^{-1}}$, the largest values being located in the lower troposphere at high and low latitude ($\approx 1.5$~kJ~kg${^{-1}}$). 
From Fig.\ref{Fig_3}, $|a_p|\approx 100$~kJ~kg${^{-1}}$ and $a_T \approx 1$ to $5$~kJ~kg${^{-1}}$.
The specific available enthalpy, $a_h$, is thus separated into a small positive reservoir, $a_T$, and a larger component, $a_p$, which has a zero global integral. 
Therefore, $|a_p| \gg a_T$ and $a_h \approx a_p$.

The further partition of $a_T$, into $a_B + a_S + a_C$ can be viewed similarly as a hierarchical separation of the positive energy reservoirs $a_B$ and $a_S$ from the component $a_C$ whose isobaric average cancels. From Figs.\ref{Fig_5} and \ref{Fig_6}, typical values of $a_B$, $a_S$ and $|a_C|$ are $0.2$~kJ~kg${^{-1}}$, $2.5$~kJ~kg${^{-1}}$ and $1$~kJ~kg${^{-1}}$, respectively.

\begin{figure}[t]
\centering
\includegraphics[width=0.6\linewidth,angle=0,clip=true]{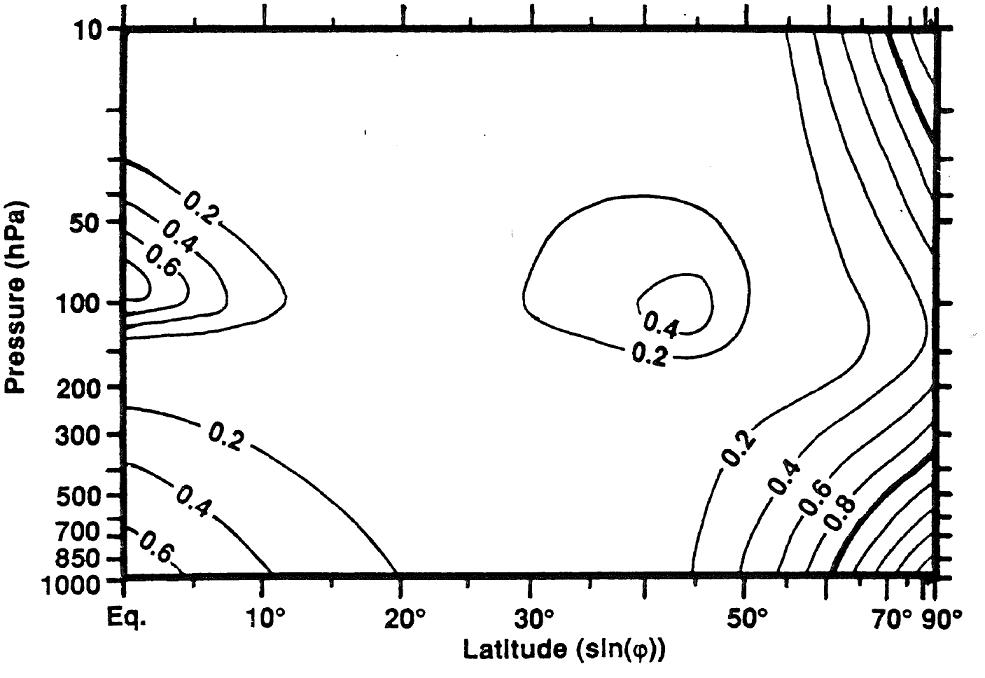}
\caption{
{\it
Distribution of the positive baroclinicity component of the available enthalpy formulation $a_B$, in
kJ~kg${}^{-1}$.
}
\label{Fig_4}}
\end{figure}

\begin{figure}[t]
\centering
\includegraphics[width=0.6\linewidth,angle=0,clip=true]{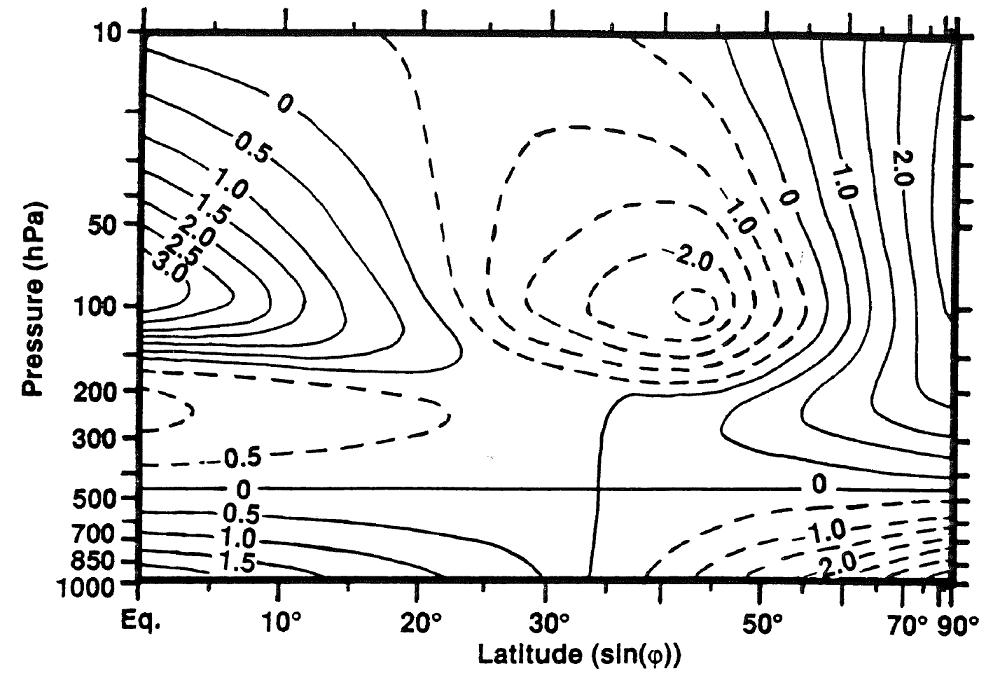}
\caption{
{\it
Distribution of the component $a_C$, in kJ~kg${}^{-1}$ (dashed lines denote negative values).
}
\label{Fig_5}}
\end{figure}

\begin{figure}[t]
\centering
\includegraphics[width=0.45\linewidth,angle=0,clip=true]{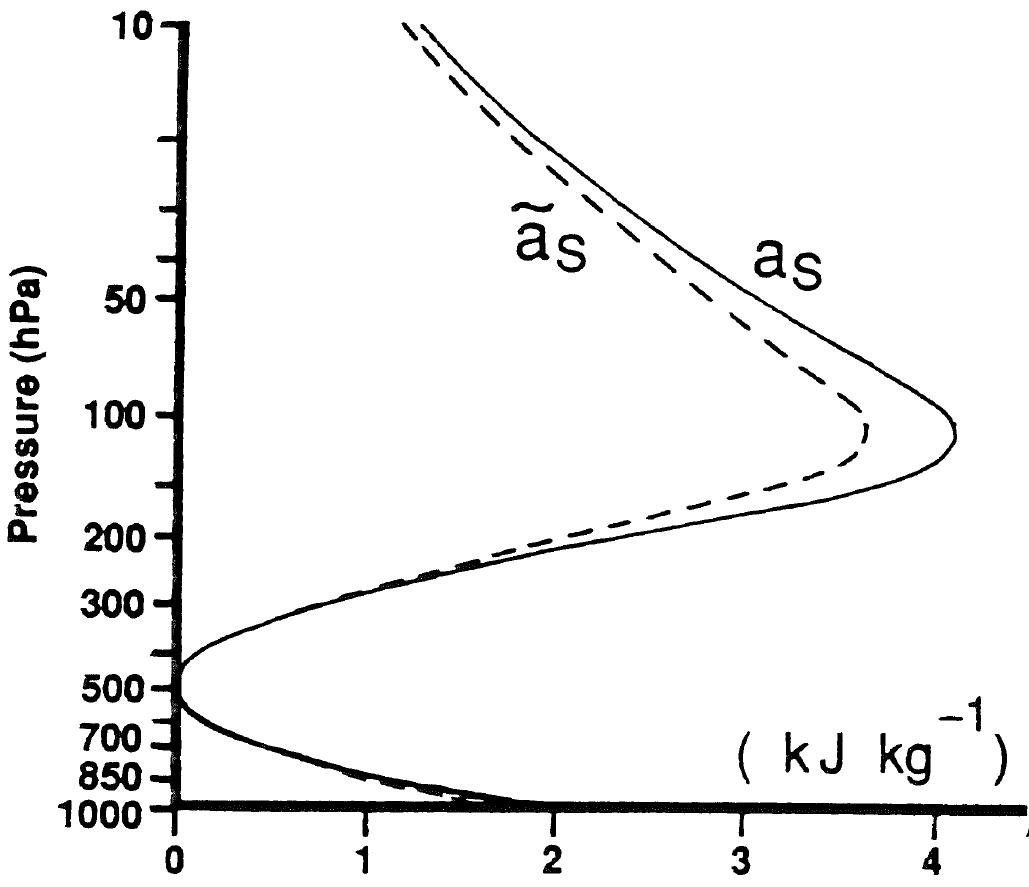}
\includegraphics[width=0.45\linewidth,angle=0,clip=true]{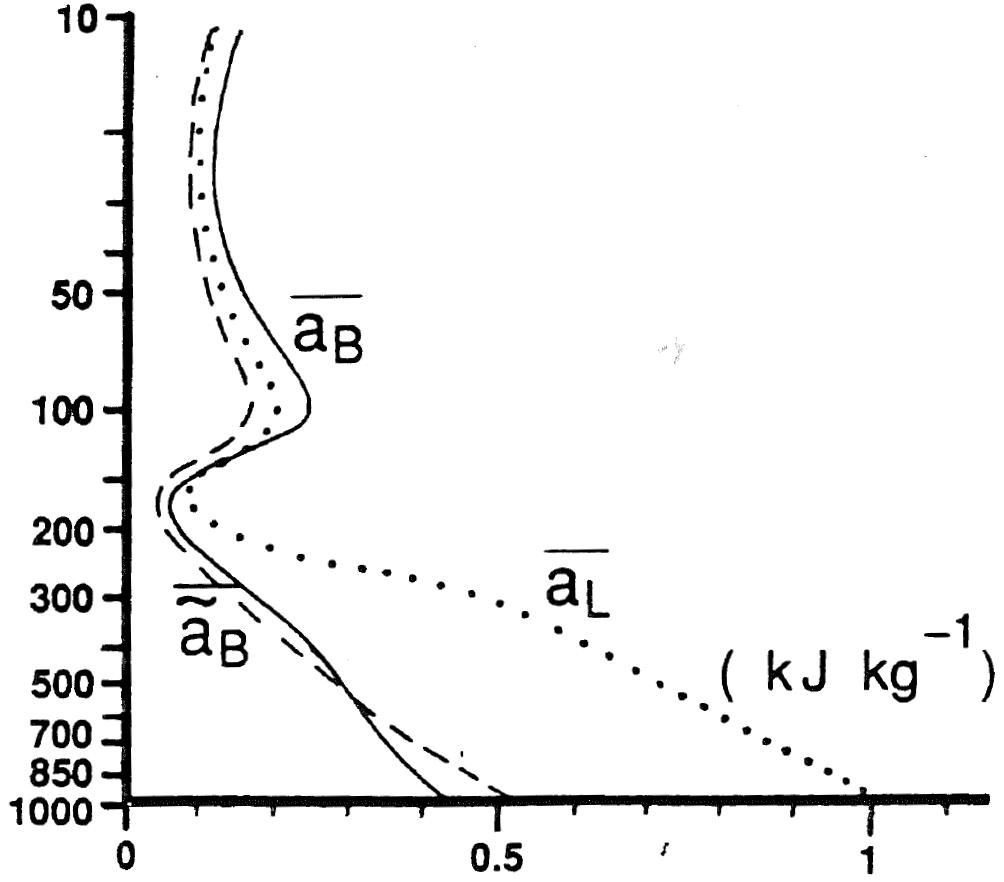}
\\ \hspace*{0cm} (a) \hspace*{9cm} (b)
\caption{
{\it
(a) Vertical profile of the static stability components. 
The available enthalpy quantity $a_S$  is shown as a solid line; $\widetilde{a}_S$ (Pearce's quantity) is shown as a dashed line. 
(b) Isobaric averages of the baroclinicity components: $\overline{a_B}$ are shown as a solid line; $\overline{\widetilde{a}_B}$ (Pearce's quantity) are shown as a dashed line.
The isobaric averages of $a_L$ (the Lorenz component) are shown by a dotted line.
Units are kJ~kg${}^{-1}$.
}
\label{Fig_6}}
\end{figure}

The spatial distributions of $a_B$ and $a_L$ are mainly controlled by the isobaric variations of temperature: their patterns are very similar (Fig.\ref{Fig_2} may be compared to Fig.\ref{Fig_4}). 
But the isobaric averages of $a_L$ and $a_B$ depicted in Fig.\ref{Fig_6}(b) show that the values of $\overline{a_B}$ are smaller (larger) than the values of $\overline{a_L}$ in the troposphere (stratophere).
More exactly, $\overline{a_B}$ is about 42\% of $\overline{a_L}$ throughout the troposphere, that is for nearly 80\% of the mass of the atmosphere and for the greatest values of $\overline{a_B}$ and $\overline{a_L}$.
Therefore, this property must apply on a global average: $A_B \approx APE/2$.
Pearce had indeed already found changes in Lorenz's global APE about twice those in his global baroclinicity component $\widetilde{A}_B$ (see Fig.6 of Pearce 1978, where $\widetilde{A}_B = \widetilde{A}_Z + \widetilde{A}_E$).

Tropospheric and stratospheric values of $\overline{a_B}$ are of the same order of magnitude
(Fig.\ref{Fig_6}(b)). 
Thus, the vertical profile of $\overline{a_B}$ is more balanced than the profile of $\overline{a_L}$, whose
large lower-tropospheric values prevail. 
The difference is due to an explicit contribution from $\overline{\sigma}$ (the stability parameter) in the definition (\ref{eq_43}) of $\overline{a_L}$, whereas in the present theory (and in Pearce's theory) all the static stability effect is taken into account by the component $a_S$. 
The vertical profile of $\overline{a_L}$ reaches extreme values in the boundary layer where static instabilities could occur.
This would make Lorenz's approximate expression inapplicable ($\overline{\sigma} \rightarrow 0 \Rightarrow a_L \rightarrow \infty$). 
In contrast, both the components $a_B$ and $a_S$ remain well defined, even if $\overline{\sigma} \leq 0$. 
Moreover, the component $a_S$ does not depend on the vertical derivative of temperature but only on the deviation from the constant profile $\overline{T} = T_r$, making the computations of $a_S$ easier than for $a_L$ in the case of a crude vertical resolution.

The difference between $a_S$ and $\widetilde{a}_S$ is solely due to the approximation ${\cal F}(X_S) = X_S/2$, which is true within about 10\%. 
Nevertheless, it must be noticed that this difference can reach $0.47$~kJ~kg${^{-1}}$ near the tropopause (Fig.\ref{Fig_6}(a)), a value which is not negligible in comparison with the values of $a_B$ at the same level: $0.25$~kJ~kg${^{-1}}$.

Therefore, the discrepancies with Pearce's formulation lead to significant numerical differences for the energy reservoirs $a_B$ and $\widetilde{a}_B$, as well as $a_S$ and $\widetilde{a}_S$ .

Finally, the specific kinetic energy has a wide range of values. 
For velocities in the range from $10$~m~s${^{-1}}$ (lower troposphere) to $50$~m~s${^{-1}}$ (jets), $e^{\star}_K$ varies from $0.05$~kJ~kg${^{-1}}$ to $1.25$~kJ~kg${^{-1}}$.

All these numerical evaluations allow us to consider the cycle (\ref{eq_37}) as energetic processes between isobaric average energy components of increasing order of magnitude:
\begin{equation}
  \left( \; \; \overline{e^{\star}_K} \; \sim \; \overline{a_B} \; \; \right)
  \; \; < \; \; 
  \left( \; \; a_S \; \sim \; | \, a_C \, | \; \; \right)
  \; \; \ll \; \; 
  | \, a_p \, |
 \: .
\label{eq_44}
\end{equation}

Pearce has demonstrated that despite the fact that $\widetilde{A}_B < \widetilde{A}_S$ (one order of magnitude), the time variations of $\widetilde{A}_B$ and $\widetilde{A}_S$ are nearly the same for the life cycle of an idealized baroclinic wave (Fig.6 of Pearce 1978, where $\widetilde{A}_Z + \widetilde{A}_E = \widetilde{A}_B$). 
Thus, one can expect the three isobaric rates of change in (\ref{eq_37}) to be also of the same order of magnitude, and the hierarchical partition (\ref{eq_44}) leads to an exact energy cycle (\ref{eq_37}) where the various terms are balanced. 
In particular, even if locally $|a_p| \gg a_S$, the isobaric flux convergence $\overline{B(a_p)}$ (see (\ref{eq_39})) remains of the same order of magnitude as $\overline{C_{(S,K)}}$ (see (\ref{eq_38})) in the energy equation for $\overline{a_S}$.

Finally, from the partitions $T = T_r + (T - T_r)$ and $T = T_r + ( \overline{T} - T_r) + (T - \overline{T} )$, the introduction of $T_r$ in section~\ref{section_Ah} and the use of $\overline{T}(p)$ in section~\ref{section_energy_components} can be interpreted as different approximations of the atmospheric thermodynamic state.
\begin{itemize}[label=,leftmargin=3mm,parsep=0.1cm,itemsep=0cm,topsep=0cm,rightmargin=2mm]
\vspace*{-1mm}
\item  i) The isothermal zero-order atmosphere at constant temperature $T_r$ (for which the pressure for the scale height is $p_r$) provides the zero-order approximation and leads to the splitting of $a_h$ into $a_p(p)$ and $a_T( T - T_r)$. 
\item  ii)  The first-order approximation is given by the vertical profile of the isobaric temperature, namely $\overline{T}(p)$, which leads to the further separation of $a_T$ into $a_C$, $a_S(\overline{T}-T_r)$ and $a_B(T-\overline{T})$.
\end{itemize}

Therefore, $a_T$ and $a_S$ are the first-order available enthalpy reservoirs, since they only depend on $\overline{T}-T_r$, or $T-\overline{T}$, which are two deviations from the zero-order term $T_r$. 
The baroclinicity component, $a_B$, depends only on a deviation from a first-order term, namely $T-\overline{T}(p)$: the baroclinicity component can be interpreted as the second-order available enthalpy component.

 \section{SUMMARY AND REMARKS.} 
\label{section_summary}

In this paper we have shown that the concept of available enthalpy, $a_h$, previously introduced in thermodynamic theory can be used to study local properties of atmospheric energetics. 
This approach appears to generalize that of Lorenz (1955,1967) and is based rather on the quasi-local approach of energetics described by Pearce (1978).

The link with the global static entropic energy ($T_0 \: \Sigma$) introduced by Dutton (1973, 1976) has not really been made, although it is mentioned in section~\ref{section_HIST_BACK} that $T_0 \: \Sigma$ already corresponds to what is called the (global) non-flow gross-work function in exergy theory.
Actually, $T_0 \: \Sigma$  is the primary concept of energy availability introduced by the founders of thermodynamics.

The local properties elucidated by the available enthalpy are the local energy cycle (\ref{eq_14}), Bernoulli's law (\ref{eq_20}) and the hydrostatic cycle (\ref{eq_29}). 
The exact global conservation law (\ref{eq_24}) generalizes the properties of Lorenz's APE discussed in section~\ref{section_global_conservation_law}, except that the global available enthalpy is zero for an isothermal atmosphere at temperature $T_r$ (the reference state of Pearce). 
The hydrostatic cycle for isobaric average energy components (\ref{eq_37}) is the local (that is for a prescribed value of pressure) generalization sought by Pearce.
This cycle has almost the same global behaviour as Lorenz's cycle (conversion, generation and dissipation terms), except that the energy reservoirs are not the same ($A_B$ + $A_S$ instead of APE). 
The differences from the formulation of Pearce are the conversion term $\overline{C_{(S,K)}}$ and the flux convergence terms related to the components $a_C$ and $a_p$ in the equation for $a_S$. 
All these new terms vanish on a global average, which is the case in the global formulations of Lorenz and Pearce.

The problem of an uneven topography has given rise to numerous studies (Taylor 1979; Koelher 1986; Boer 1989), but none of them has really succeeded in improving the formulation of Lorenz. 
This problem is easily solved in this study since the flat earth assumption is not used to derive the locaI approach. 
It must be noticed, however, that in the cycle (\ref{eq_37}), the time derivatives are taken at constant pressure. Thus, special attention must be given to possible points of intersection of lower isobaric layers with topography when permuting the isobaric averages operator with the isobaric changes in time. 
However, upper isobaric layers are not concerned at all and (\ref{eq_37}) is always valid, whereas an uneven topography modifies these upper levels in Lorenz's reference state (by rearranging the whole Lorenz's reference state due to any local change of entropy and $\theta$ close to the surface).

The guiding principle chosen by Pearce in formulating a concept of meteorological available energy agrees with the presentation adopted above for the definition of available enthalpy. 
Pearce gave up the concept of a reference state which would possess a maximum kinetic energy and which could actually be reached through a redistribution of the mass.
Rather, he chose as a starting point prescribed relations between sources, sinks and changes in a suitable available energy. 

The available enthalpy approach is different from the available energy concept of Pearce in that it is a general thermodynamic local state function which has proved to be significant in order to derive what Baehr (see Haywood 1974) described as the part of (thermodynamic) energy that can be transformed into any other form of energy. 
But this exergy approach was not so easy to introduce in atmospheric energetics since the temperature of the thermostat (which is the central concern of the exergy theories) must be replaced by the definition (\ref{eq_18}) of a mere numerical value $T_r$ related to a space-time average over the whole fluid. 

It is a little disappointing that this meteorological available enthalpy cannot be rigorously introduced from the general thermodynamic  theory\footnote{\color{blue} 
It is however worth noting that the Kullback-Liebher information can be taken as the starting point to define exergy functions. If entropy is defined by the Shannon's formula $S = -\:\sum_i P_i \ln(P_i)$ in terms of the microstates $P_i$, the Kullback-Liebher divergence (also called ``relative entropy'' or ``information gain'' or ``information divergence''...) is equal to $D_{KL}= +\:\sum_i P_i \ln(P_i/P_{i0})$ where the $P_{i0}$ represent the microstates of ``ambient conditions'' associated with the macroscopic temperature $T_0$ and pressure $p_0$.
See
\url{http://en.wikipedia.org/wiki/Kullback-Leibler_divergence}.
It is shown for instance in Karlsson (1990) that the macroscopic value associated with  $D_{KL}$ is the non-flow exergy!
}: 
it has been arbitrarily defined in the present paper (although, as mentioned in section~\ref{section_local_energy_law}, the author had rediscovered the function $a_h$ and all its local and global physical properties, starting from meteorological considerations, before becoming aware of the concept of  exergy)\footnote{\color{blue} 
Soon after my own (re-)discovery of the function $(h-h_r)-T_r\:(s-s_r)$ in 1989, I have searched in thermodynamic courses whether or not this function could correspond to any known function?
Alas, I have then understood that Gouy's function and Gouy-Stodola theorem already exist in my student thermodynamic book, in French!
I have then scrutinized old books and articles available at the Polytechnic School, and soon discovered that {\it Exergy\/} is the modern name and concept associated with availability in energy or enthalpy.
Then, I have not tried to hide these old facts and to reinvent a new name (like the ``APE'' of Lorenz or the ``static entropic energy'' of Dutton).}.
However, the concept of potential change in total entropy is a typical meteorological definition somewhat similar to that of potential temperature (see the footnote ${}^{\mbox{\scriptsize\ref{footnote_Gibbs}}}$ and  Section~\ref{section_potential_entropy}).

One can recognize an analogy with a historical thermodynamic problem: the interpretation of the ``heat function'' or ``heat content'' also called ``heat power at constant pressure'' which was closely related to steady-flow and constant pressure conditions in the early definitions of what is nowadays called enthalpy. 
The enthalpy function is simply considered as a mere thermodynamic state function. 
Similarly, the function $a_h$ leads to numerous results which have been derived in this study, even if a priori it should not have been applied to atmospheric energetics (namely if one would be concerned with searching for {\it attainable \/} reference states).

One might consider the function $a_h$ as a mere thermodynamic state function which can be applied independently of the exergy theory. 
In particular it must be stressed that what is sometimes called ``reference state'' in the present study should have been more accurately called ``the reference thermodynamical constants $T_r$ and $p_r$''. 
This is the main difference from the theory of Dutton and it is rather in agreement with Pearce's proposal.

To be more convinced by the soundness of using the concept of exergy instead of more ordinary thermodynamics as defined by Margules and Lorenz, one may notice that exergy-like quantities are joint properties of two principles of thermodynamics (thermodynamic potentials), and that they take into account the possibility of occurrence of natural irreversible thermodynamic processes (the links between irreversibilities and exergy theory have been widely studied since Jouguet 1907). 
Indeed, quantities such as Carnot's factors (efficiency factors) arise in exergy theory from irreversibilities occurring during heat flow between the thermostat and the fluid. 
Lorenz (1967) has interpreted differently the meteorological efficiency factor. 
The static entropic energy of Dutton, $T_0 \: \Sigma$, is also a thermodynamic potential. 
An important use of the thermodynamic potentials (Gibbs or Helmholtz functions for instance) is to obtain simple criteria for the natural sense of processes, and to discover thermodynamic equilibrium states: this has been done by Dutton (1973, 1976). 
The quantity $T_0 \: \Sigma$ which is used in Livezey and Dutton (1976) is related to the potential 
$E_i + P_0 \: V - T_0 \: S$ 
(see also the discussion of Landau and Lifchitz 1958, sections 19 to 21, relating to this potential).

But even if $a_h$ is also a thermodynamic potential-like function (at least mathematically since $h$ and $-T_r \: S$ both appear and are related to the first and second law of thermodynamics, respectively) it seems that it has not been studied in this sense, either in the present paper or elsewhere as far as the author is aware. The reason is that $a_h$ is only a semi-convex function, i.e. only the component $a_T(T)$ possesses convexity properties about $T = T_r$, owing to the mathematical form of ${\cal F}(X)$, whereas $a_p(p)$ appears to not have, at first sight, even a prescribed sign\footnote{\color{blue} 
It is mentioned in footnote ${}^{\mbox{\tiny\ref{footnote_ap_Margules}}}$ that vertical integrals of $a_p$ corresponds to Margules' function which is quadratic in surface pressure...
Therefore, the ``problem'' of no ``definite sign of $a_p$'' is not a real problem!}.
On the contrary, $T_0 \: \Sigma$ is positive and doubly convex as is shown by Dutton (1973, 1976) in meteorology or similarly by Bejan (1987) in thermodynamics\footnote{\color{blue} 
It is not certain that this is a clear advantage, because the flow exergy formulation $a_p = R\:\ln(p/p_r)$ leads to an interesting global integral which involves quadratic functions in surface pressure (and not directly depends on $(p-p_r)^2$ as suggested by Dutton's non-flow exergy formulation).}.


On the other hand, the concept of specific available enthalpy has interesting local properties, as shown in the present paper, and several applications can be expected for this local energetics. 
It is for instance suitable for studying the energetics of stratospheric or monsoon circulation as well as baroclinic wave events\footnote{\color{blue} Application to baroclinic waves (real and idealized ones) are described in my PhD thesis (1994) and in two QJRMS papers published in 2003.}. 
Indeed, each of these possible applications involves various vertical structures (such as upper-level jets, lower-level monsoons and wind shears) and strong boundary flux. 
Previous global meteorological available energies (Lorenz, Pearce) or static entropic energy (Dutton) were unable to differentiate these phenomena, whereas local available enthalpy energetics allows their exact study using the cycles (\ref{eq_14}), (\ref{eq_29}) or (\ref{eq_37}).

One can imagine further generalizations of the present theory and immediately think of an algebraic definition of the moist energetics where humidity is considered as an independent energy component (the treatment of Lorenz 1978, 1979 only led to numerical evaluations). 
This generalization to the moist atmosphere will be presented in a future paper in which the concept of specific moist available enthalpy will be introduced and associated with the thermodynamics of Glansdorff and Prigogine (1971).
The thermodynamical essergy (Haywood 1974; Evans 1980) is an approach which already generalizes the exergy theory to the case of chemical reactions (and indeed there are exchanges between different water phases in the moist atmosphere which may be viewed as chemical reactions). 
But the future specific moist available enthalpy will be somewhat different from essergy.\footnote{\color{blue} 
It is rather based on the approach of de Groot and Mazur (1962,1984). This moist-air generalization is published in Marquet (1993) and in my PhD Thesis (1994, in French).}

Another possible application is to the oceanographic concept of available energy (see e.g. Oort {\it et al.\/} 1989) which could be improved since one can expect the salinity to be taken into account in a new way starting with exergy-like concepts. Evans in 1969 (quoted in Haywood 1974) and Livezey and Dutton (1976) have already investigated this problem.\footnote{\color{blue} This problem is still open in January 2014.}

\vspace{5mm}
\noindent{\large\bf Acknowledgements}
\vspace{2mm}

This work has been performed at the Laboratory on Dynamic Meteorology (LMD) at the Polytechnic School  (Palaiseau, France) during a two-year post-engineering-school stay of the French school of meteorology (Ecole Nationale de la M\'et\'erologie, France).
I thank especially Dr. D. L. Cadet for his constant support during this study as well as Dr. R. Sadourny for his many helpful suggestions and comments. I am also very grateful to J. F. Mahfouf and A. Lasserre-Bigorry who read early versions of this paper.
Thanks are also due to R. P. Pearce and J. A. Dutton for their helpful comments and suggestions.
Computations were made at the C.N.R.S. computing centre (C.I.R.C.E.).

\vspace{8mm}
\noindent
{\large\bf Appendix A. List of symbols.}
             \label{appendixSymbol}
\renewcommand{\theequation}{A.\arabic{equation}}
  \renewcommand{\thefigure}{A.\arabic{figure}}
   \renewcommand{\thetable}{A.\arabic{table}}
      \setcounter{equation}{0}
        \setcounter{figure}{0}
         \setcounter{table}{0}
\vspace{1mm}
\hrule

\vspace*{-1mm}
\begin{tabbing}
 -----------------------\=  ------------------------------------------------------------------------- --\= \kill
 $p$, $T$, $V$ \> Pressure, temperature and volume\\
 $p_0$, $T_0$ \> Two outer reference values (general exergy theory)\\
 $p_r$, $T_r$ \> Two reference values (present available enthalpy theory)\\
 $c_{p}$  \> Specific heat of dry air at constant pressure \>($1004$~J~K${}^{-1}$~kg${}^{-1}$) \\
 $c_{v}$  \> Specific heat of dry air at constant volume   \>($717$~J~K${}^{-1}$~kg${}^{-1}$) \\
 $R$      \> Gas constant og dry-air \>($287$~J~K${}^{-1}$~kg${}^{-1}$) \\
 $\kappa=R/c_{p}=2/7$ \>\\
 $H$, $h$ \> Total and specific enthalpy \\
 $S$, $s$ \> Total and specific entropy \\
 $G=H - T \: S$ \> Gibbs function (``free enthalpy'') \\
 $p_{00}=1000$~hPa \> Standard constant pressure \\
 $\theta = T\:(p_{00}/p)^{\kappa}$   \> Potential temperature \\
 $\rho$   \> Density \\
 $\alpha = 1 / \rho $ \> Specific volume \\
 $T_m(t)$ \> A uniform temperature (Pearce, 1978) \\
 $Q = T \: ds/dt$ \> Diabatic heating rate per unit mass \\
 $g$ \> Magnitude of gravity ($9.81$~m~s${}^{-2}$) \\
 $\lambda$ \> Longitude \\
 $\varphi$ \> Latitude \\
 $z$ \>  Upward distance (zero at sea level) \\
 $w = dz/dt$ \>  $z$-component of velocity\\
 $\omega = dp/dt$ \>  Vertical wind component in isobaric coordinates\\
 $\eta$ \>  A dummy scalar\\
 $r$ \>  Radius of the Earth\\
 $d\Sigma$ \> Element of horizontal area \\
 $dm$ \>  Element of mass\\
 $d \tau$ \>  Element of volume\\
 $\phi = g \: z$ \> Geopotential \\
-----------------------\=  ---------------------------------------------- --\= \kill
 $E_i$, $e_i$ \>  Internal energy \> (total and specific value)\\
 $E_K$, $e_k$ \>  Kinetic energy \> (total and specific value)\\
 $E_G$, $e_g$ \>  Gravitational potential  energy \> (total and specific value)\\
 $A_h$, $a_h$ \>  Available enthalpy \> (total and specific value)\\
 $A_p$, $a_p$ \>  Pressure component of $A_h$ \> (total and specific value)\\
 $A_T$, $a_T$ \>  Temperature component of $A_h$ \> (total and specific value)\\
 $A_B$, $a_B$ \>  Baroclinicity component of $A_h$ \> (total and specific value)\\
 $A_S$, $a_S$ \>  Static stability component of $A_h$ \> (total and specific value)\\
 $A_C$, $a_C$ \>  Complementary component of $A_h$ \> (total and specific value)\\
 $A_N$, $a_N$ \>  Anergy component of $A_h$ \> (total and specific value)\\
 $A_L$, $a_L$ \>  Lorenz's available potential energy \> (total and specific value)\\
 $B(\eta)$ \>    Isobaric divergence of the scalar $\eta$\\
 $N_h$, $N_{\theta}$ \> Efficiency factors \\
 $P_{\theta}$ \> Isentropic average pressure \\
 $\overline{\sigma}$ \> Lorenz stability parameter (in isobaric average)\\
 $\Delta S^0$ \> Potential change in total entropy \\
 $G_{\alpha}$ \> Generation of $\alpha$-energy component \\
 $C_{(\alpha,\beta)}$ \> Conversion from $\alpha$-energy to $\beta$-energy component \\
 $D_K$ \> Frictional dissipation of (kinetic) energy \\
 $ds$, $q$ \> Elementary change in entropy, heat quantity (both in section~\ref{section_potential_entropy}) \\
 ${\cal F}$ \>  The (non-flow exergy) function ${\cal F}(X) \: =  \: X \: - \: \ln(1+X)$, for $X>-1$ \\
 ${\cal K}$ \>  A constant in Bernoulli's theorem\\
 ${\cal S}$ \>  Horizontal surface of an atmospheric (limited space) domain\\
 ${\cal M}$, ${\cal D}$  \>  Mass and volume integrating domain of the atmosphere\\
 $M$ \>  Value of the mass of the atmosphere \\
 $T_0 \: \Sigma $ \> Static entropic energy (Dutton, 1973, 1976) \\
 $t$ \> Time \\
 $\Delta t = t_2 - t_1$ \> A time interval \\
 $d/dt$ \> Total, material, Lagrangian or natural time derivative operator \\
 $\partial/\partial t$ \> Local (also Eulerian) time derivative operator \\
 $.$ \> Scalar product of two vectors \\
 $\vec{v}$ \> Wind vector (three-dimensional) \\
 $\vec{u}$ \> Horizontal wind vector (two-dimensional) \\
 $\vec{F}$ \> Specific frictional force (three-dimensional) \\
 $\overrightarrow{\nabla}$ \> Gradient operator (three-dimensional) \\
 $\overrightarrow{\nabla}_{\!p}$ \> Gradient operator at constant pressure (two-dimensional and horizontal) \\
-----------------------\=  ------------------------------------------------------------------------- --\= \kill
\end{tabbing}
\vspace*{-6mm}
Subscripts
\vspace*{-4mm}
\begin{tabbing}
 -----------------------\=  ------------------------------------------------------------------------- --\= \kill
 $r$  \> Denote reference values (related to the arbitrary state $T_r$, $p_r$) \\
 $0$  \> Denote the reference state of exergy theory (outer medium, state  $T_0$ and $p_0$) \\
      \> or equilibrium state in Dutton studies
\end{tabbing}
\vspace*{-3mm}
Superscripts
\vspace*{-4mm}
\begin{tabbing}
 -----------------------\=  ------------------------------------------------------------------------- --\= \kill
 ${\ast}$  \> Denote hydrostatic term\\
 $'$       \> Denote deviation from isobaric average \\
 $\circ$  \> For the potential change in total entropy ($\Delta S^0$ )
\end{tabbing}
\vspace*{-1mm}
Overbars
\vspace*{-2mm}
\begin{tabbing}
 -----------------------\=  ------------------------------------------------------------------------- --\= \kill
 $\overline{\mbox{(...)}}$  \> Isobaric average\\
 $\widehat{(...)}$          \> Global average\\
 $\,\widetilde{(..)}$       \> Denote Pearce's terms
\end{tabbing}

\newpage 

\vspace{4mm}
\noindent
{\large\bf Appendix B. A review of Exergy formulae.}
             \label{appendix_B}
\renewcommand{\theequation}{B.\arabic{equation}}
  \renewcommand{\thefigure}{B.\arabic{figure}}
   \renewcommand{\thetable}{B.\arabic{table}}
      \setcounter{equation}{0}
        \setcounter{figure}{0}
         \setcounter{table}{0}
\vspace{1mm}
\hrule
\vspace{3mm}

Since the emergence of concepts related to availability in energy, various names and terminologies have been used in thermodynamic literature. 
It seems helpful to review at least those used in the different papers or books quoted in the present article. 
Hard copies of some of the old papers are included in this arXiv version, but they were not included in Marquet (1991).

The papers of Gibbs (1873a,b) quoted in this paper use a geometric approach, with almost no explicit analytic formulas for available energies (see however the footnote ${}^{\mbox{\scriptsize\ref{footnote_Gibbs}}}$ and Section~\ref{section_potential_entropy}, where explicit formulae are given).

Notations have been sometimes brought up to date to give a homogeneous presentation. 
In this appendix, $U$ is the internal energy, $H$ or $h$ the enthalpy, $T$ the temperature, $S$ or $s$ the entropy, $E_c$ the kinetic energy, $P$ the potential energy, ${\mu}_n$ and $N_n$ the chemical potential and the concentration of the species $n$. 
Subscripts $i = (0, 1, 2)$ denote thermodynamic states of pressure $p_i$ and temperature $T_i$.

\vspace{5mm}
{\large\bf  B-1. \underline{Thermodynamics literature}}
\vspace{2mm}

\begin{figure}[t]
\centering
\includegraphics[width=0.98\linewidth,angle=0,clip=true]{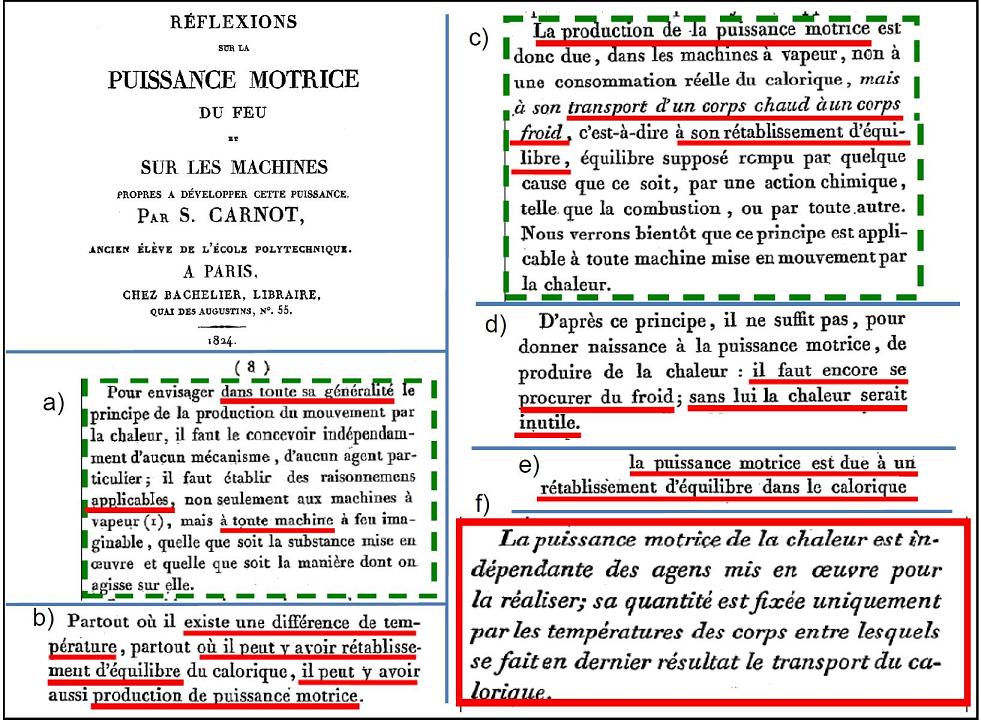}
\caption{
{\it
``R\'eflexions sur la puissance motrice du feu et sur les machines propres \`a d\'evelopper cette puissance''
(``Reflections on the Motive Power of Heat and on machines fitted to develop that power'').
Nicolas L\'eonard Sadi Carnot (1924).
}
\label{Fig_7}}
\end{figure}

{\bf $\bullet \;$ Nicolas L\'eonard Sadi Carnot (1824). ``Reflections on the Motive Power of Heat and on machines fitted to develop that power''. See Fig.\ref{Fig_7}\footnote{\color{blue} 
This paper published in 1824 was not described in Marquet (1991).}}

\begin{figure}[t]
\centering
\includegraphics[width=0.98\linewidth,angle=0,clip=true]{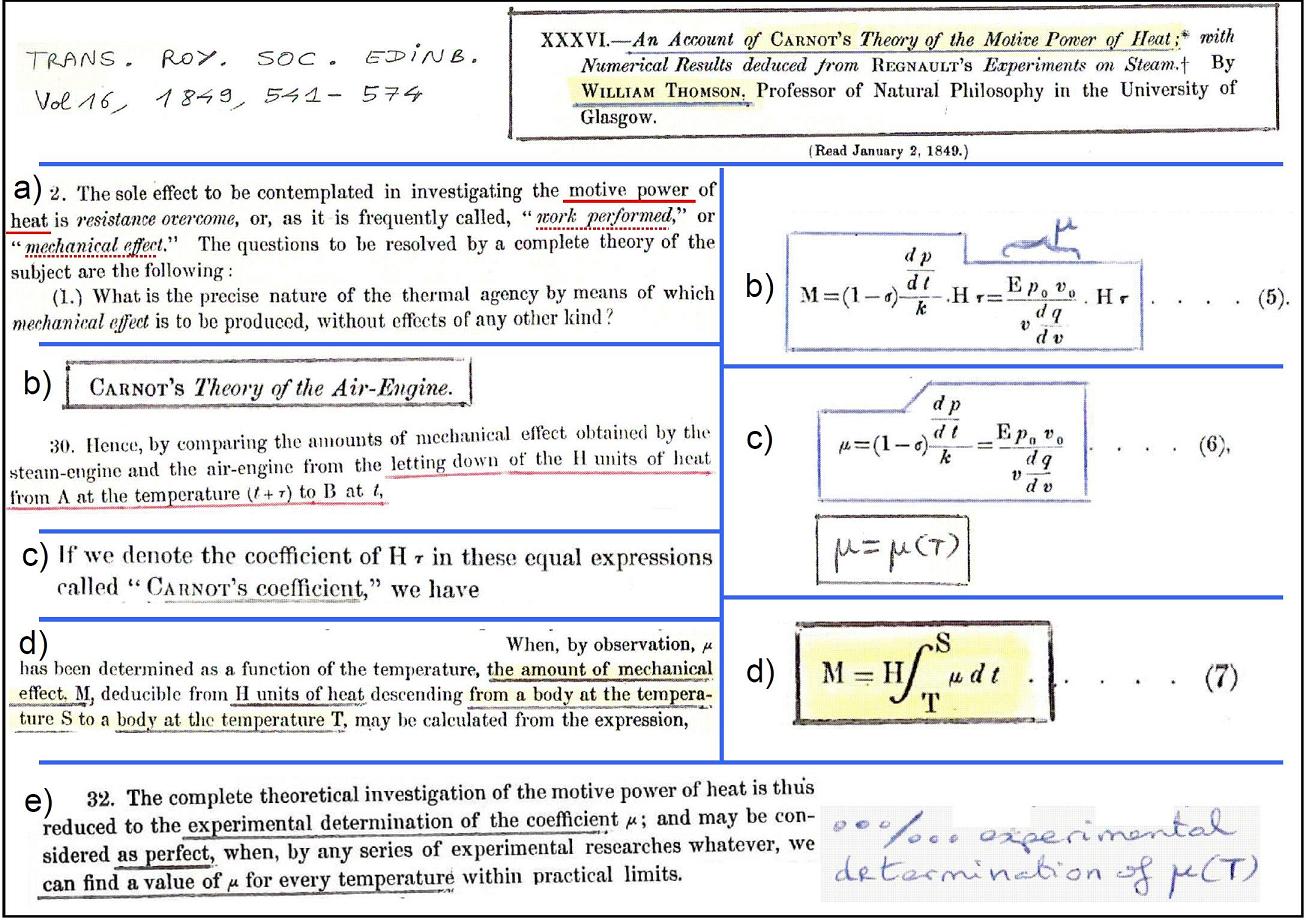}
\caption{
{\it
``An account of Carnot's theory of Motive power of heat''.
William Thomson (1849).
}
\label{Fig_8}}
\end{figure}

The text of Carnot did not contain mathematic formula.
It is however clear that the ``Motive Power of Heat'' is not an energy.
It corresponds to what is called ``Available Energy'' or ``Exergy'' nowadays!
An English translation by R. H. Thurston of the version published in the ``Anales scientifique de l'\'Ecole Normale Sup\'erieure'' (ii. series, t.1, 1872) is available in the url: \url{http://www3.nd.edu/~powers/ame.20231/carnot1897.pdf} (Wiley \& Sons, 1897, digitized by Google).
\begin{itemize}[label=,leftmargin=3mm,parsep=0cm,itemsep=0.1cm,topsep=0cm,rightmargin=2mm]
\vspace*{-1mm}
\item  a) 
In order to consider in the most general way the principle of the production of motion by heat, it must be considered independently of any mechanism or any particular agent. 
It is necessary to establish principles applicable not only to steam-engines* but to all imaginable heat-engines, whatever the working substance and whatever the method by which it is operated.
(* We distinguish here the steam-engine from the heat-engine in general. 
The latter may make use of any agent whatever, of the vapor of water or of any other, to develop the motive power of heat.)
\item b)
Wherever there exists a difference of temperature, wherever it has been possible for the equilibrium of the caloric to be re-established, it is possible to have also the production of ``impelling power'' (i.e. of ``Motive power'').
\item c)
The production of motive power is then due in steam-engines not to an actual consumption of caloric, but to its transportation from a warm body to a cold body, that is, to its re-establishment of equilibrium -- an equilibrium considered as destroyed by any cause whatever, by chemical action such as combustion, or by any other. 
We shall see shortly that this principle is applicable to any machine set in motion by heat.
\item d)
According to this principle, the production of heat alone is not sufficient to give birth to the impelling (Motive)  power: it is necessary that there should also be cold; without it, the heat would be useless.
\item e)
the motive power is due to a re-estabishment of equilibrium in the caloric
\item f)
{\it The motive power of heat is independent of the agents employed to realize it; its quantity is fixed solely by the temperatures of the bodies between which is effected, finally, the transfer of the caloric\/.} 
This is the fundamental result used in next studies: $\mu(T) = 1/T$ only depends on (absolute) temperature, and not of the kind of working substance or steam-engine!
\end{itemize}


\vspace{2mm}
{\bf $\bullet\;$ William Thomson / Lord Kelvin (1849, p. 556, Eq.7): an account of works of Sadi Carnot (1824) and Henri Victor Regnault (1847). See Fig.\ref{Fig_8}\footnote{\color{blue} This first paper published in 1849 was not mentioned in Marquet (1991).}}
\begin{eqnarray}
 M \; = \; \delta Q\; \int_{T_0}^T \mu(T') \: dT' 
  & \hspace{10mm} \mbox{\underline{Amount of mechanical effect}} \nonumber
\end{eqnarray}
$M$ was called ``the amount of mechanical effect deductible from a unit of heat $\delta Q$ descending from a body at the temperature $T$ to a body at the temperature $T_0$''.
The ``Carnot's coefficient'' $\mu(T)$ must only depend on the (absolute) temperature $T$ (this is the main result of Carnot, 1824).
Thomson mentioned that $\mu(T)$ was still to be determined from experimental determination.
The notations was different in Thomson (1849): $H$ for $\delta Q$, $S$ for $T$, $T$ for $T_0$, $t$ for the dummy integrating temperature $T'$.
If $\mu(T) = 1/T$, then $M = \delta Q \; \ln(T/T_0)$.
In modern terminology, this is the change in entropy associated with a change in temperature from $T_0$ to $T$ (though the concept of entropy was not invented by Clausius (1865), nor the concept of absolute temperature by Thomson!)


\begin{figure}[t]
\centering
\includegraphics[width=0.98\linewidth,angle=0,clip=true]{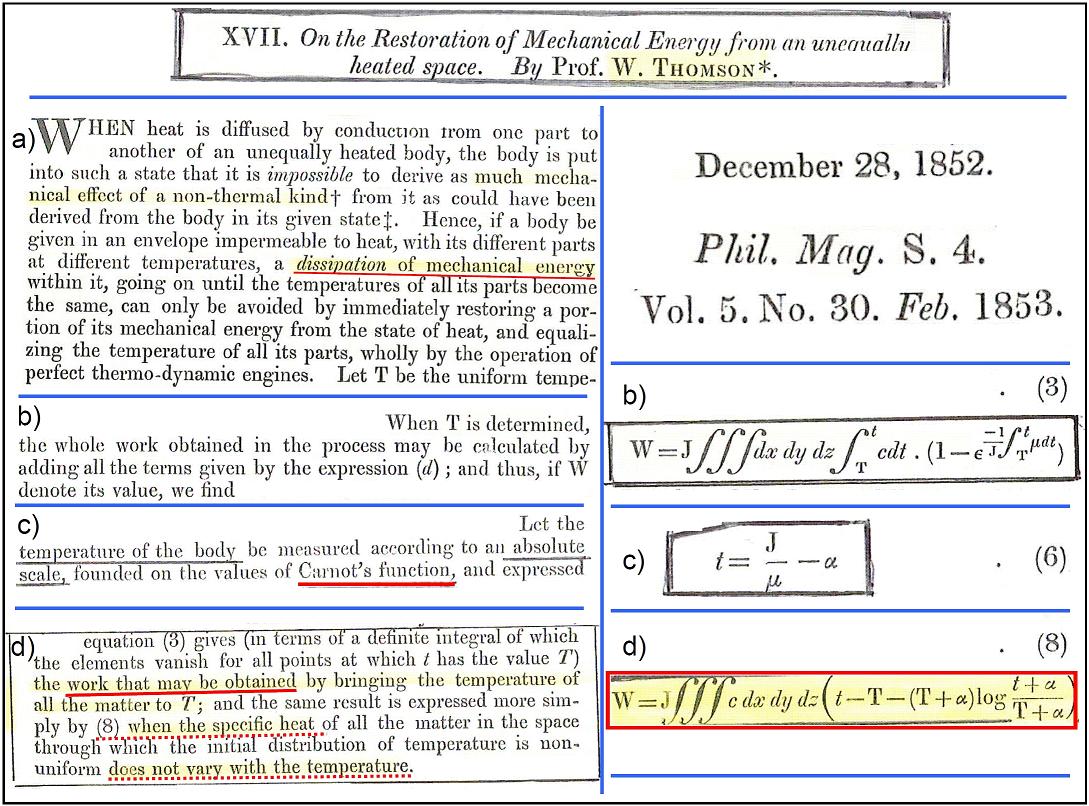}
\caption{
{\it
``On the restoration of mechanical energy from an unequally heated space''.
William Thomson (1853).
}
\label{Fig_9}}
\end{figure}

\begin{figure}[t]
\centering
\includegraphics[width=0.95\linewidth,angle=0,clip=true]{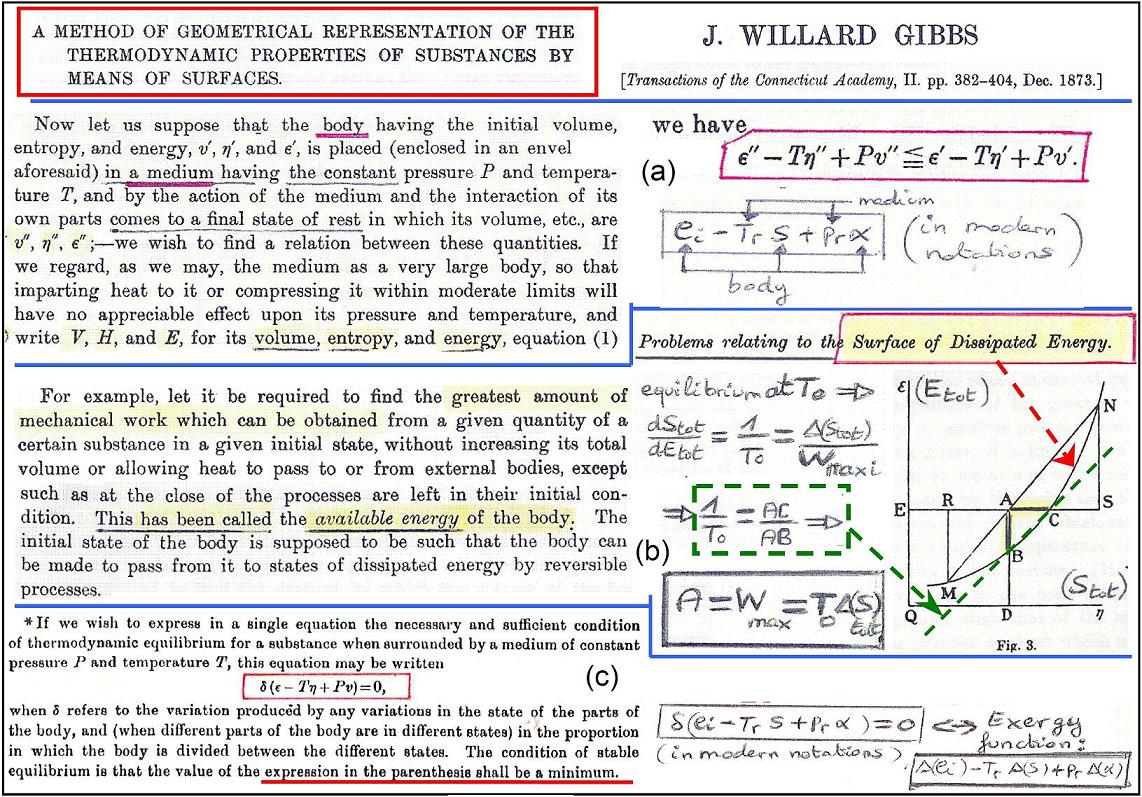}\\
\includegraphics[width=0.95\linewidth,angle=0,clip=true]{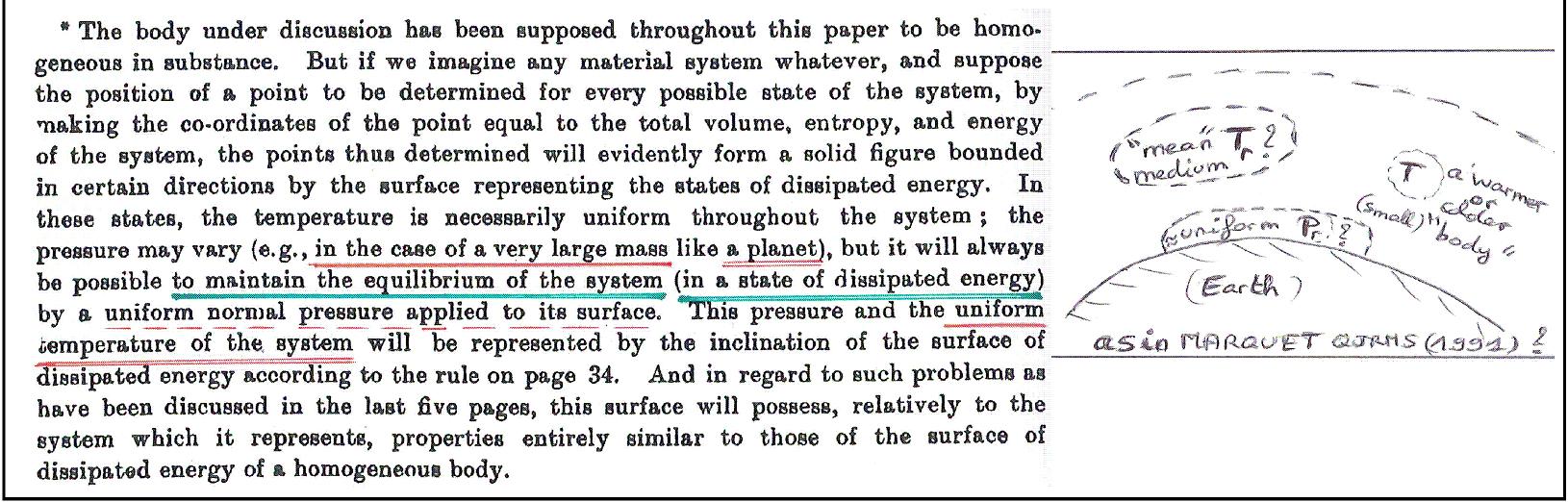}
\caption{
{\it
``A method of geometrical representation of the thermodynamic properties of substance by means of surface''.
Josiah Willard Gibbs (1873b).
}
\label{Fig_10}}
\end{figure}

\vspace{2mm}
{\bf $\bullet \;$ WilliamThomson / Lord Kelvin: ``On the restoration of mechanical energy from an unequally heated space'' (1853). See Fig.\ref{Fig_9}}
\begin{eqnarray}
\hbox{Eq.3 (p.104):} \; \; \; \; 
  W \; = \;
 \iiint c_p \; \left[ \: 
 \int_{T_0}^T  
          \left( \: 
          1 - \frac{T_0}{T'}
         \: \right)\: dT'
    \; \; \right]
    dx \: dy \: dz \: .
 & \hspace{5mm} \mbox{\underline{Available work}} \nonumber
\end{eqnarray}
\begin{eqnarray}
\hbox{Eq.8 (p.105):} \; \; \; \; 
  W \; = \;
 \iiint c_p \; \left[ \: 
          (T-T_0) - T_0 \: \ln(T/T_0)
             \: \right] \: 
    dx \: dy \: dz  \: .
  & \hspace{5mm} \mbox{\underline{Available work}} \nonumber
\end{eqnarray}
The notations was different in Thomson (1853): $t+\alpha$ for $T$, $T+\alpha$ for $T_0$, a factor $J$ before the integral.
In modern notations: $\alpha=0$ (this leads to the definition of absolute temperature $T$) and $J=1$ (the mechanical equivalent of the thermal unit, in Joule unit nowadays).
Carnot's coefficient was set to $\mu=J/(t+\alpha)$, which is equal to $1/T$ in modern notations.
It is worth noting that the function $(T-T_0) - T_0 \: \ln(T/T_0)$ is exactly equal to ${\cal F}(T/T_0-1)$, i.e. to the  exergy functions appearing in the non-flow or flowing exergy functions (like the available temperature component $a_T$ of the available enthalpy $a_h$).
This result published in 1853 by Thomson was obtained whereas the concept of entropy was not yet isolated by Clausius (1865)! (the concept of absolute temperature was just  defined some years before by Thomson himself, in 1848).


\vspace{2mm}
{\bf $\bullet \;$ Josiah Willard Gibbs  ``A method of geometrical representation of the thermodynamic properties of substance by means of surface'' (1873, Volume~I of the Collected works,  1928). See Fig.\ref{Fig_10}}\footnote{\color{blue} The content of this  paper was mentioned in the main text, but not in Appendix~B of Marquet (1991).}
\begin{eqnarray}
     (U - U_0) \: - \: T_0 \: (S - S_0) \: + \: p_0 \: (V-V_0) \; \geq \; 0 \; .
  & \hspace{5mm} \mbox{\underline{called non-flow exergy nowadays, p.40}} \nonumber
\end{eqnarray}
\begin{eqnarray}
     W_{max} \; = \;  A \; = \;
    T_0 \; \Delta S_{tot} \; .
  & \hspace{10mm} \mbox{\underline{Capacity for entropy, p.51}} \nonumber
\end{eqnarray}
Only the first formulas was  explicit in the paper of Gibbs (though expressed with old notations).
The second one is expressed in terms of the change in total entropy $\Delta S_{tot}$ represented by the horizontal distance AC in in Fig.3 of Gibbs (1973).
The maximum change in total energy (called available energy $A=W_{max}$ nowadays) is equal to the vertical distance AB in the same Fig.3 and the two distances are linked via the slope of the ``surface of dissipated energy'' equal to $\Delta S_{tot} / W_{max}  = \:$AC/AB$\:=1/T_0$,  leading to $A = T_0 \; \Delta S_{tot}$.

The last important remark depicted in the bottom case of Fig.\ref{Fig_10}  is made in the last page~54 of Gibbs (1873b).
It can be understood as a prophetic view of the application to the global atmosphere, with the constant temperature $T_r$ imposing the inclination of the surface of dissipated energy via AC/AB$\:=1/T_r$.
This is the generalization to the non-homogeneous atmosphere retained in my paper (Marquet 1991), where $a_h$ can indeed be defined by $T_r \: \Delta S_{tot} = (h-h_r) - T_r \: (s - s_r)$.
This explains why the introduction of $T_r$ as a mean (average) value is relevant, with no attempt to define a real  and attainable ``reference state'' (Section~\ref{section_potential_entropy}).


\begin{figure}[t]
\centering
\includegraphics[width=0.95\linewidth,angle=0,clip=true]{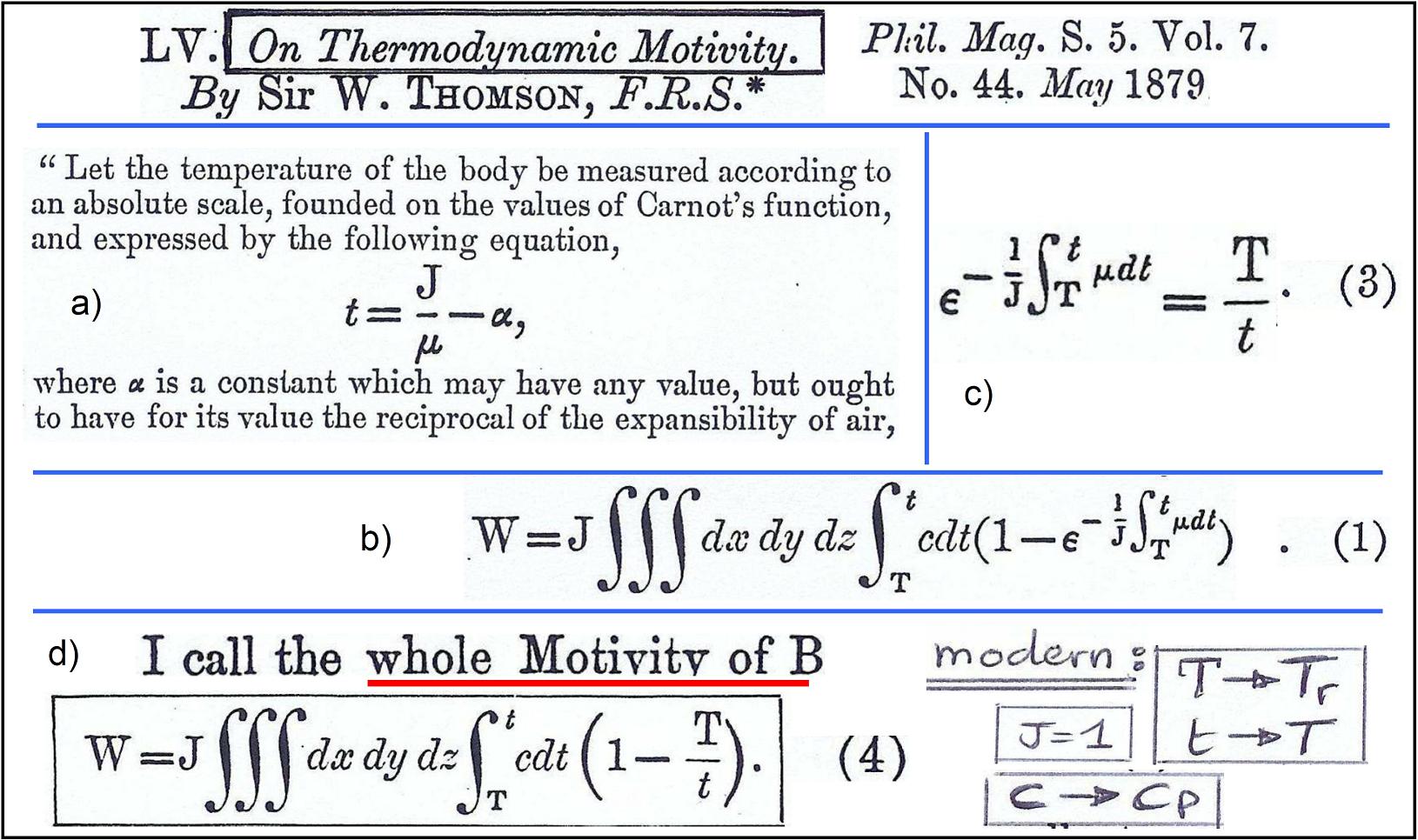}
\caption{
{\it
``On thermodynamic Motivity''.  William Thomson (1879).
}
\label{Fig_11}}
\end{figure}

\vspace{2mm}
{\bf $\bullet \;$ William Thomson / Lord Kelvin: ``On thermodynamic Motivity'' (1879, p.350, Eq.4). See Fig.\ref{Fig_11}.}
\begin{eqnarray}
  W \; = \; \iiint
   \left[ \; \;
 \int_{T_0}^T  c_p \; 
          \left( \: 
          1 - \frac{T_0}{T'}
         \: \right)\: dT'
    \; \; \right]
    dx \: dy \: dz \; .
  & \hspace{5mm} \mbox{\underline{Motivity}} \nonumber
\end{eqnarray}
It is based on  Eq.3 published in 1853 with the ``Carnot's coefficient'' $\mu(T)$ set to the reciprocal of the (absolute) temperature $1/T$.
The {\em Motivity} is defined for a given distribution of temperature at $T$ surrounded by other matter all at a common temperature $T_0$.
$W$ is the work obtainable from this given distribution of temperature at $T(x,y,z)$ by means of perfect thermodynamic engines interacting with the matter at $T_0 = \;$constant.
The available work published in Eq.8 of Thomson (1853) corresponds to this Motivity with a constant value of $c_p(T,p)$. 


\begin{figure}[t]
\centering
\includegraphics[width=0.95\linewidth,angle=0,clip=true]{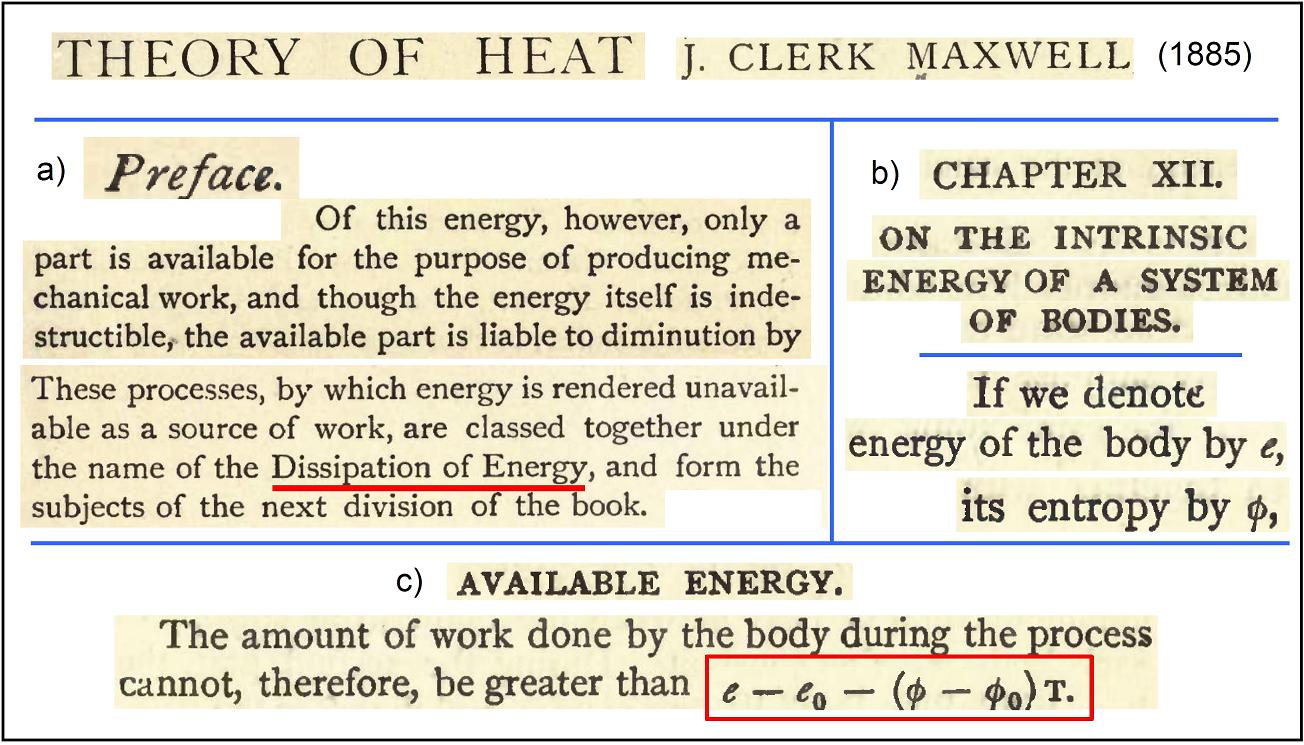}
\caption{
{\it
``Theory of heat''.  James Clerk Maxwell (1885).
}
\label{Fig_12}}
\end{figure}

\vspace{2mm}
{\bf $\bullet \;$ James Clerk Maxwell: ``Theory of heat'' (1885, Chapter XII in the 8th of a first edition in 1870). See Fig.\ref{Fig_12}}\footnote{\color{blue} 
This entry was labeled ``1871'' in Marquet (1991).}.
\begin{eqnarray}
     (U-U_0) \: - \: T_0 \; (S-S_0) \; .
  & \hspace{10mm} \mbox{\underline{Available energy}} \nonumber
\end{eqnarray}
The first editions (1870-71) of the book of Maxwell did not provide this explicit formula and the available energy was erroneously called ``entropy'' by Maxwell in first editions of the book (still in the third one in 1872), being influenced by the Scottish mathematical physicist P. G. Tait and  differently from the way Rudolf Clausius (1865) has defined the modern version of this concept.
According to Gibbs (1973) the above formula was written in the next editions of ``Theory of heat'', where the energy ($U$) and the entropy ($S$) was, for instance, denoted by $e$ and $\phi$ in the edition of 1902 (with corrections and additions made in 1891 by Lord Rayleigh), leading to the formula $e - e_0 - (\phi - {\phi}_0)\: T$, where $T$ is equivalent to the constant value $T_0$ in modern exergy theories.
This notation $\phi$ for the entropy gave rise to the concept of ``Tephigram'', i.e. the thermodynamic diagram where (dry-air) entropy and temperature  ($T$--$\, \phi$ diagram) are straight lines in the  plot.


\vspace{2mm}
{\bf $\bullet \;$ Louis Georges Gouy (1889): ``Sur l'\'energie utilisable'', Journal de Physique, in French (``On the available energy'', in the Journal of Physics).}
\begin{eqnarray}
     d\,{\cal E} \; = \; 
   d\, (U\: - \: T_0 \; S) 
   \: + \: d\,W \; ,
  & \hspace{5mm} \mbox{\underline{Energie utilisable} / \underline{Available energy}} \nonumber
\end{eqnarray}
where ${\cal E}$ is the available energy, $U$ the total energy and $S$ the entropy.


\vspace{2mm}
{\bf $\bullet \;$ Aurel Sotola (1898): ``Die Kreisprozesse der Gasmaschine'', Zeitschrift des Vereines Deutscher Ingenieure, in German (The cyclic processes of gas engine).}

A series of paper of Aurel Stodola published in 1903  has been translated in French in 1906 by E. Hahn (``Les turbines \`a vapeur'' / ``The steam machines''.).
\begin{eqnarray}
   A \; = \; 
     (H_2-H_1) \: - \: T_0 \; (S_2-S_1)
     \; = \;
    T_0 \: \Delta\, S_{tot} \; .
  & \hspace{5mm} \mbox{\underline{Energie utilisable} / \underline{Available work}} \nonumber
\end{eqnarray}
It was about the first time that the available energy $A$  was defined for flowing fluids in terms of the difference of enthalpies $H_2-H_1$ of a fluid flowing through a device (a surrounding medium at $T_0$), this replacing the difference in internal energy $U_2-U_1$  for non-flowing fluids.
Stodola also explained that the maximum available work $A$  is equal to the product of $T_0$ by the increase of entropy of all matter involved by the transformation of the flowing fluid passing from state $1$ to state $2$.
This corresponds to the definition of Gibbs ($T_0 \: \Delta\, S_{tot}$).
The Gouy--Stodola theorem was the first entry for my discovery of  exergy principles.


\vspace{2mm}
{\bf $\bullet \;$ Emile Jouguet  (1907), Georges Darrieus (1931), in French.}
\begin{eqnarray}
     (U_2-U_1) \: - \: T_0 \; (S_2-S_1) \; .
  & \hspace{5mm} \mbox{\underline{Energie utilisable} / \underline{Available energy}} \nonumber \\
     (H_2-H_1) \: - \: T_0 \; (S_2-S_1) \; .
  & \hspace{5mm} \mbox{\underline{Enthalpie utilisable} / \underline{Available enthalpy}} \nonumber
\end{eqnarray}


\begin{figure}[t]
\centering
\includegraphics[width=0.95\linewidth,angle=0,clip=true]{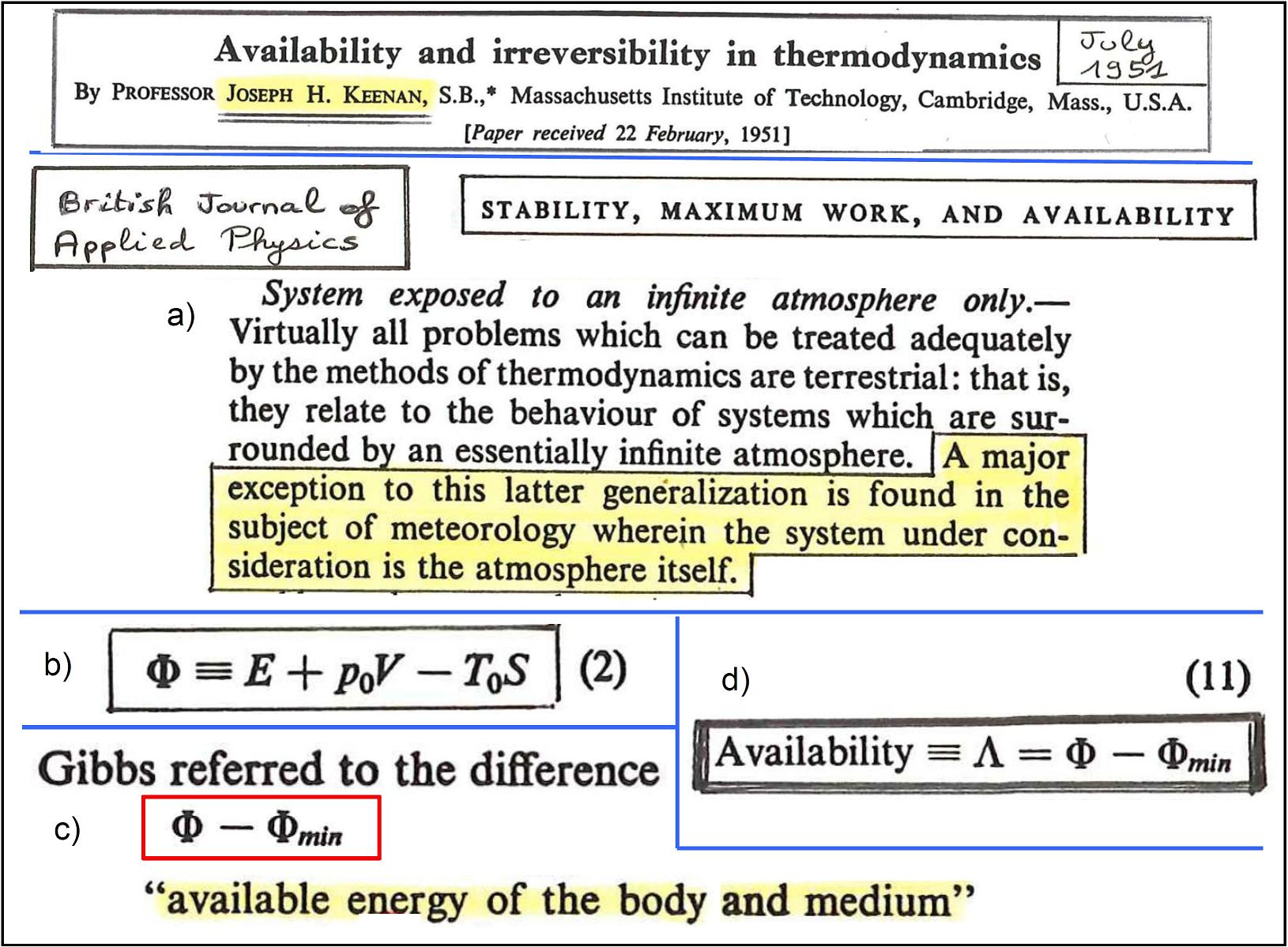}
\caption{
{\it
``Availability and irreversibility in thermodynamics''.  Joseph H. Keenan (1951).
}
\label{Fig_13}}
\end{figure}

\vspace{2mm}
{\bf $\bullet \;$  Joseph H. Keenan (1932, 1951). See Fig.\ref{Fig_13}}.
\begin{eqnarray}
    \hbox{(1932)} \; \; \; \;   b  \; = \; h - \: T_0 \; s 
  & \hspace{0mm} \mbox{(not named)} \nonumber 
\end{eqnarray}
\begin{eqnarray}
    \hbox{(1932)}: \; \; \; \;  
        b_2 \: -  \:  b_1  \; = \; ( h_2 - h_1 ) - \: T_0 \; ( s_2 - s_1 )
    \; =
  & \hspace{0mm} \mbox{\underline{Change in  Availability}} \nonumber 
\end{eqnarray}
\begin{eqnarray}
    \hbox{Eq.2,  p.184 (1951):} \; \;  \; \;  \Phi \; =  \;   U  \: + \: p_0 \; V \: - \: T_0 \; S \; .
  &  \nonumber 
\end{eqnarray}
\begin{eqnarray}
    \hbox{Eq.11,  p.185 (1951):} \; \; \; \;  \Lambda \; = \;   \Phi   \: - \: {\Phi}_{min}  \; .
  & \hspace{2mm} \mbox{\underline{Availablility}} \nonumber 
\end{eqnarray}
\begin{eqnarray}
    \hbox{Eq.14,  p.185 (1951):} \; \; \; \;  \Delta \Lambda \; = \;   \Delta  \Phi  \; .
  & \hspace{2mm} \mbox{\underline{Increase in Availablility}} \nonumber 
\end{eqnarray}

Keenan made reference to studies of Thomson, Gibbs, Maxwell and Darrieus.
However, notations $b$, $\Phi$ or $\Lambda$ are unusual for exergy quantities.
The remark of Keenan (1951) which predict that ``A major exception (...) is found in the subject  of meteorology'' is no so relevant, since the global available enthalpy is indeed a relevant generalization of Lorenz's concept!


\begin{figure}[t]
\centering
\includegraphics[width=0.95\linewidth,angle=0,clip=true]{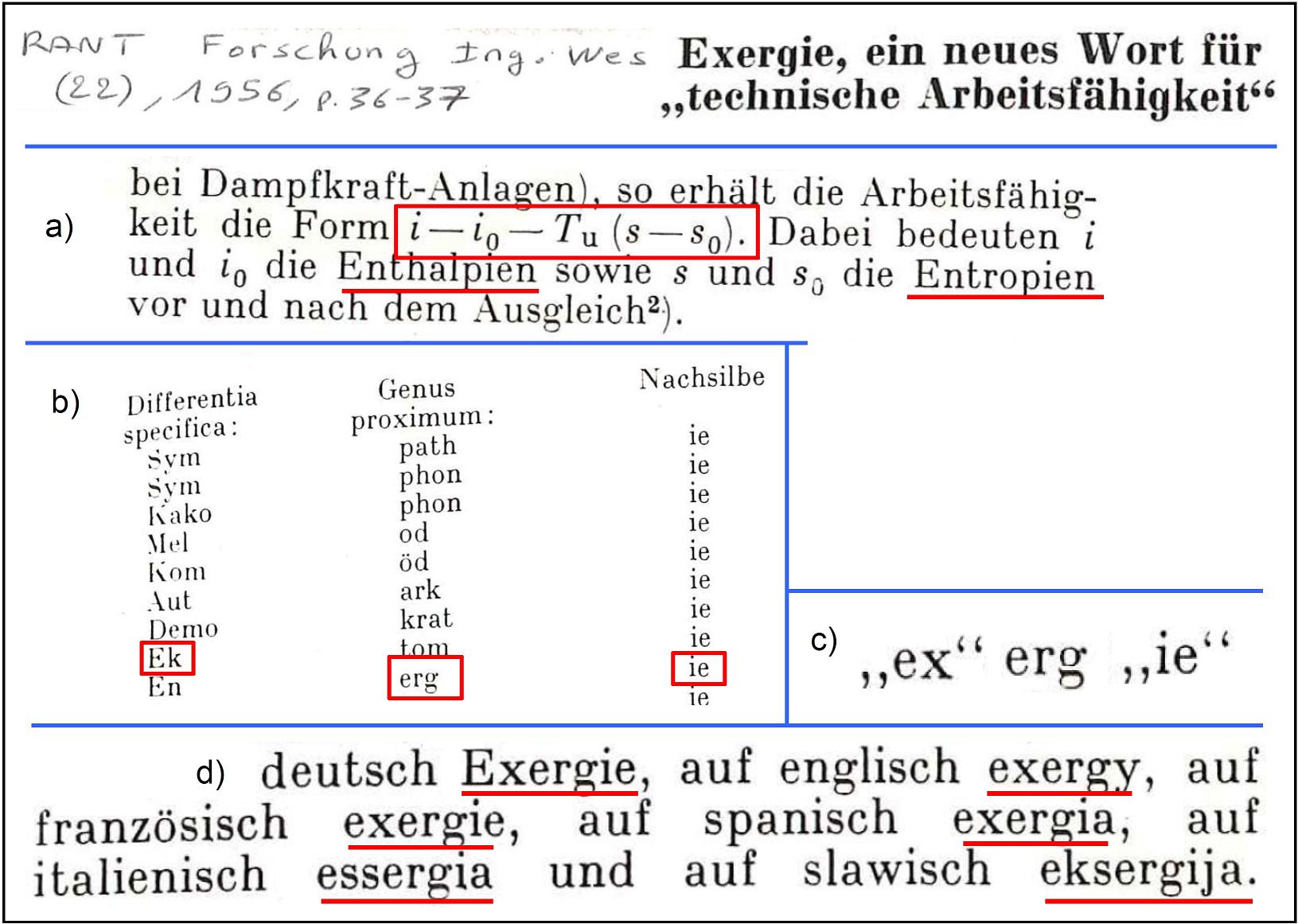}
\caption{
{\it
``Exergie, ein neues Wort f\"{u}r Technische Arbeitsf\"{a}higkeit''.  Zoran Rant (1956).
(Exergy, a new name for technical ``capacity for work'' or ``available work'').
}
\label{Fig_14}}
\end{figure}

\vspace{2mm}
{\bf $\bullet \;$  Zoran Rant  (1956). See Fig.\ref{Fig_14}.}
\begin{eqnarray}
     (h - h_0)   \: - \: T_0 \; (s-s_0) \; .
  & \hspace{5mm} \mbox{\underline{Exergy} (here Rant coined the name)} \nonumber 
\end{eqnarray}

A possible English translation of the German text is suggested hereafter.

{\it 
When a body changes from a given (thermodynamic) state to another, the maximum amount of work is obtained if  transformation is reversible. 
For each of these state, that is to say for each ``energy'', one can associate  a ``capacity for technical work'' or more simply a ``capacity for work'' (or ``available work''?). 
The way to compute this capacity for work depends, however, on  nature of  energy. 
In most of cases (in particular for kinetic or electric energies) the capacity for work coincides with energy itself.
}

{\it 
Transformations of state, where the bodies are not initially in equilibrium (though in contact) with the environment at temperature $T_e$ and pressure $p_e$, are of a special importance for the definition of the efficiency values. 
Values of $T_e$ and $p_e$ represent constant, reference values. 
Let us consider a thermodynamic system where a certain amount of heat $Q$ enters and leave this system. 
If $T$ is the outgoing temperature, then the capacity for work is equal to $Q \: (T- T_e )/ T$. 
If, moreover, a certain amount of matter enters the system (as observed in all the turbines, heat engines or steam machines), then the capacity for work is equal to $i - i_0 - T_e \: (s - s_0)$.
Here, $i$ and $i_0$ are the enthalpies and $s$ and $s_0$ are the entropies, respectively before and after the process leading to the equilibrium state characterized by $i_0$ and $s_0$
}

{\it 
It is possible to imagine that some unknown forms of energy could exist for which the capacity for work ought to be computed differently. 
In the literature, the notion of capacity for work is clearly associated with the special form of energy dealing with``matter'' ; it is the reason why it is too strongly associated with this kind of energy (of matter).
By analyzing various processes by which different kind of matter or energies are transformed, and by examining the
associated capacity for work, it is possible to extend this notion (of capacity of work), in a very logical way. 
Similarly, P. Gra{\ss}mann has proposed the concept of ``technical powers'', which could be a generalization of the other ones.
}

{\it
It is worthwhile to note: 
that we have first defined the concept of capacity of work and only afterward seek for the way to compute it; 
that the computation of the capacity of work depends on the kind of energy from which it is derived and the
formulas are different depending on each kind of energy; 
that the use of the enthalpy is only associated with the capacity of work due to exchanges of matters; 
and that the capacity of work represents a true work (in fact the maximum of it, really available from the system via reversible processes).
}

{\it
The capacity of technical work has became such an important concept, either in technical domain or in (fundamental) physics, that one should follow the proposition of R. Planck to find a new name for it, with the idea to give to it a general recognition in the vocabulary of the physics community. 
The proposition of U. Grigull to call the capacity for work ``Ecthalpie''  is not acceptable to the opinion of the present author (Z. Rant). 
Indeed, a consequence of the previous paragraphs is that the capacity of work does not represent a ``thalpie'' (heat) and neither an ``ecthalpie'' (a delivering of heat). 
It represents a true work. 
We will show that it is possible to apply a general method to determine the new name for ``capacity of work'', by following the expected demand for the invention of a new world to be accepted internationally (namely in several languages). 
The result is that the ``capacity of work'' must be called ``\underline{Exergie}'' (in German), which is the most suitable for international applications.
}

{\it
In order to be suitable for international purposes, and for a sake of equity, a new word must not correspond to any of the modern languages. 
It must take its inspiration from the classical languages, as the Latin or the Greek. 
 Moreover, the five following constraints ought to be fulfilled.
}

{\it \noindent
1) Expression must be short enough to be able to build derived expressions. 
It is especially important for the Latin or Slav languages, for which it is difficult to agglomerate existing words to form another (new) word (contrary to the Germanic ones, and especially for the German, for which it is easier to do so). 
In Latin and Slav languages, it is mainly possible to add prefixes or suffixes to existing short root-names. 
It is reason why names composed of several root-names will not be considered here.
}

{\it \noindent
2) The searched expression must be understandable by itself (namely to contain the definition of the concept) in order to avoid the need of extra explanations. 
This definition is made of two parts: the ``genus proximum'' (the closest root-name) and the ``differentia specifica'' (the root that makes the definition different from all other ones). 
The modern (living) languages require, however, another element, in order to make the final expression a name belonging to the considered language: this is obtained with the aid of suffixes which are specific to the considered language (--t\'e, --isme, --ie etc). 
Therefore, the searched expression must contain at least three syllabus (and may be more). 
Here is a list of such derived expressions coming from the Greek (see Fig.\ref{Fig_14}).
(Note: resulting names in the previous table are the ones published in the original paper of Z. Rant.
They generate German or French words but the change of the suffix ``ie'' into ``y'' would generate the equivalent English words.)
}

{\it \noindent
3) The resulting expression must be closed to the set of preexisting expressions corresponding to related or similar concepts. 
The thermodynamical quantity (``capacity for work'') for which we search for a new expression is very close to the (usual thermodynamical) state variables. 
For some of them, we (already) use expressions coming from the Greek: (internal) energy, enthalpy, entropy. 
The new expression (for ``capacity of work'') must verify the same requirements.
}

{\it \noindent
4) If the etymology of the new expression must be close to the etymology of the other similar expressions (i.e. like energy, enthalpy or entropy), it must be sufficiently different to avoid any misunderstanding with the (already existing) other expressions
}

{\it  \noindent
5) The new expression must ``sound in a good manner''...
}

{\it
In order to verify all the requirements 1) to 5), it arises that ``ie'' is the more relevant suffix (for German, French or Spanish languages, corresponding to ``y'' for the English one). 
As for the ``genus proxinum'', it must correspond to the name ``work'' and the root-name must be the corresponding Greek name ``erg(on)''. 
The last choice concerns the prefix, corresponding to the specific particularity or to the ``differentia specifica''. 
It is clear that the ``capacity for work'' is also a ``work that can extracted from a given system''. 
``Extracted from'' writes ``ec'' in Greek before a consonant and it writes ``ex'' before a vowel.
}

{\it
Therefore, the new name for ``capacity for work'' must write ``\underline{ex-erg-ie}'', that is to say \underline{exergie}. 
This new name verifies all the requirements1) to 5) mentioned above. 
In particular the letter ``x'' makes the name ``exergy'' clearly different from its parent one ``energy'', and any mistaken between the two words is avoided. 
The main expression ``\underline{exergie}'' can be translated into the Germanic, Latin or Slav languages, leading to ``\underline{Exergie}'' in German, ``\underline{exergy}'' in English, ``\underline{exergie}'' in French, ``\underline{exergia}'' in Spanish, ``\underline{essergia}'' in Italian and ``\underline{eksergija}'' in Slav.
}

{\it
Written by Zoran RANT at Lublijana on the 17th of November 1955.
}


\vspace{2mm}
{\bf $\bullet \;$  Raymond Marchal  (1956), in French.}
\begin{eqnarray}
     (U_2-U_1)  \: + \: p_0 \; (V_2-V_1) \: - \: T_0 \; (S_2-S_1) \; .
  & \hspace{5mm} \mbox{\underline{Energie utilisable} / \underline{Available energy}} \nonumber 
\end{eqnarray}
\begin{eqnarray}
     (H_2-H_1) \: - \: T_0 \; (S_2-S_1) \; .
  & \hspace{5mm} \mbox{\underline{Enthalpie utilisable} / \underline{Available energy with flowing}} \nonumber 
\end{eqnarray}


\vspace{2mm}
{\bf $\bullet \;$  Andr\'e Martinot Lagarde (1971), in French.}
\begin{eqnarray}
     U \: - \: T_0 \; S \; .
  & \hspace{10mm} \mbox{\underline{Energie utilisable} / \underline{Available energy}} \nonumber 
\end{eqnarray}
\begin{eqnarray}
     H \: - \: T_0 \; S \; .
  & \hspace{5mm} \mbox{\underline{Enthalpie utilisable} / \underline{Available enthalpy}} \nonumber 
\end{eqnarray}


\vspace{2mm}
{\bf $\bullet \;$  R. W. Haywood (1974).}
\begin{eqnarray}
     A^{\ast} \; = \;  U \: - \: T_0 \; S \; .
  & \hspace{15mm} \mbox{\underline{Non-flow gross-work function}} \nonumber 
\end{eqnarray}
\begin{eqnarray}
     A \; = \;  U \: + \: p_0 \; V \: - \: T_0 \; S \; .
  & \hspace{4mm} \mbox{\underline{Non-flow availability function}} \nonumber
\end{eqnarray}
\begin{eqnarray}
     B \; = \;  H  \: - \: T_0 \; S \; .
  & \hspace{18mm} \mbox{\underline{Non-flow availability function}} \nonumber 
\end{eqnarray}
\begin{eqnarray}
     A_1 \: -  \:  A_0 \; .
  & \hspace{24mm} \mbox{\underline{Non-flow exergy in state $1$}} \nonumber 
\end{eqnarray}
\begin{eqnarray}
     B_1 \: -  \:  B_0 \; .
  & \hspace{20mm} \mbox{\underline{Steady-flow exergy in state $1$}} \nonumber 
\end{eqnarray}
\begin{eqnarray}
     ( H \: -  \:  H_0) \: - \:  T_0 \; ( S \: -  \:  S_0) \; .
  & \hspace{6mm} \mbox{\underline{Exergy}} \nonumber 
\end{eqnarray}
\begin{eqnarray}
     H_0 \: + \:  T_0 \; ( H \: -  \:  H_0) \; .
  & \hspace{18mm} \mbox{\underline{Anergy}} \nonumber 
\end{eqnarray}


\vspace{2mm}
{\bf $\bullet \;$  Robert Evans (1980).}
\begin{eqnarray}
     U \: + \: p_0 \; V \: - \: T_0 \; S \: - \: \sum_n \: {\mu}_{0n} \: N_n \; .
  & \hspace{35mm} \mbox{\underline{Essergy}}  \nonumber 
\end{eqnarray}
\begin{eqnarray}
     ( H \: -  \:  H_0) \: - \:  T_0 \; ( S \: -  \:  S_0) \: + \: E_c  \: + \: E_p \; .
  & \hspace{5mm} \mbox{\underline{Exergy, useful energy, or Flow-availability}}  \nonumber 
\end{eqnarray}


\vspace{2mm}
{\bf $\bullet \;$  Lucien Borel (1987).}
\begin{eqnarray}
    J \; = \; 
       U \: + \: p_0 \; V \: - \: T_0 \; S \; .
  & \hspace{15mm} \mbox{\underline{Coenergy}}  \nonumber 
\end{eqnarray}
\begin{eqnarray}
    \widehat{J} \; = \; 
       ( H_2 \: - \: H_1 )   \: + \: p_0 \; ( V_2 \: - \: V_1 ) 
                                      \: - \: T_0 \; ( S_2 \: - \: S_1 ) \; .
  & \hspace{5mm} \mbox{\underline{Over-coenergy}}  \nonumber 
\end{eqnarray}
\begin{eqnarray}
    K \; = \; 
       H \: - \: T_0 \; S
  & \hspace{35mm} \mbox{\underline{Coenthalpy}}  \nonumber 
\end{eqnarray}
\begin{eqnarray}
    \widehat{K} \; = \; 
       ( H_2 \: - \: H_1 )  \: - \: T_0 \; ( S_2 \: - \: S_1 )   \; .
  & \hspace{15mm} \mbox{\underline{Over-coenthalpy}}  \nonumber 
\end{eqnarray}
\begin{eqnarray}
    J_{cz} \; = \;  J \: + \: E_c \: + \: P \; .
  & \hspace{6mm} \mbox{\underline{Total coenergy} \hspace{3mm}
               ($\Rightarrow  \: \widehat{J}_{cz}$: \underline{Total over-coenergy})}  \nonumber 
\end{eqnarray}
\begin{eqnarray}
    K_{cz} \; = \;  K \: + \: E_c \: + \: P \; .
  & \hspace{3mm} \mbox{\underline{Total coenthalpy} \hspace{3mm}
               ($\Rightarrow  \: \widehat{K}_{cz}$: \underline{Total over-coenthalpy})}  \nonumber 
\end{eqnarray}


\vspace{2mm}
{\bf $\bullet \;$  Michel Feidt (1987).}
\begin{eqnarray}
   ( H \: -  \:  H_0) \: - \:  T_0 \; ( S \: -  \:  S_0) \; .
  & \hspace{6mm} \mbox{\underline{Physical exergy}} \nonumber 
\end{eqnarray}
\begin{eqnarray}
   E_x  \: = \:  H  \: - \:  T_0 \;  S  \; .
  & \hspace{16mm} \mbox{\underline{Exergy}} \nonumber 
\end{eqnarray}
\begin{eqnarray}
  \hspace{28mm}  A_n  \: = \: E_{tot}  \: - \: E_x  \; .
  & \hspace{6mm} \mbox{$E_{tot}$: \underline{Total energy}; $A_n$: \underline{Anergy}} \nonumber 
\end{eqnarray}


\vspace{2mm}
{\bf $\bullet \;$  Adrian Bejan (1987).}
\begin{eqnarray}
       U \: + \: p_0 \; V \: - \: T_0 \; S \; .
  & \hspace{15mm} \mbox{\underline{Non-flow availbility}}  \nonumber 
\end{eqnarray}
\begin{eqnarray}
       H  \: - \: T_0 \; S \: + \: E_c \: + \: P \; .
  & \hspace{15mm} \mbox{\underline{Flow availbility}}  \nonumber 
\end{eqnarray}
\begin{eqnarray}
     (U-U_0)  \: + \: p_0 \; (V-V_0) \: - \: T_0 \; (S-S_0) \; .
  & \hspace{5mm} \mbox{\underline{Non-flow exergy}} \nonumber 
\end{eqnarray}
\begin{eqnarray}
     (H-H_0) \: - \: T_0 \; (S-S_0) \: + \: (E_c + P) \: - \: (E_{c0} + P_0) \; .
  & \hspace{5mm} \mbox{\underline{Flow exergy}} \nonumber 
\end{eqnarray}
\begin{eqnarray}
       U \: + \: p_0 \; V \: - \: T_0 \; S  \: - \: \sum_n \: {\mu}_{0n} \: N_n \; .
  & \hspace{15mm} \mbox{\underline{Total non-flow exergy}}  \nonumber 
\end{eqnarray}

\vspace{5mm}
{\large\bf  B-2. \underline{Meteorological literature}}
\vspace{2mm}

\vspace{2mm}
{\bf $\bullet \;$  John A. Dutton  (1973, 1976).}
\begin{eqnarray}
        T_0 \: \Sigma  \: = \:  (U-U_0)  \: - \: T_0 \; (S-S_0) \; .
  & \hspace{15mm} \mbox{\underline{Static entropic energy}}  \nonumber 
\end{eqnarray}


\vspace{2mm}
{\bf $\bullet \;$  Robert E. Livezey and John A. Dutton  (1976).}
\begin{eqnarray}
        T_0 \: \Sigma  \: = \:  (U-U_0)  \: + \: p_0 \; (V-V_0) \: - \: T_0 \; (S-S_0) \; .
  & \hspace{5mm} \mbox{\underline{Static entropic energy}}  \nonumber 
\end{eqnarray}
\begin{eqnarray}
        T_0 \: \Sigma  & = \:  (U-U_0)  \: + \: p_0 \; (V-V_0) \: - \: T_0 \; (S-S_0)
                                  \: - \: {\mu}_{0s} \: (N_s - N_{0s})  \; .  \nonumber \\
  & \mbox{(in this last formula, $N_s$ and  $N_{0s}$  are two salinities}  \nonumber 
\end{eqnarray}


\vspace{2mm}
{\bf $\bullet \;$  Robert P. Pearce  (1978).}
\begin{eqnarray}
        \frac{dA}{dt}  \: = \:  \frac{d( \, H - T_0\: S \, )}{dt}
  & \hspace{1mm} \mbox{$\Rightarrow A$: \underline{Available energy}}  \nonumber 
\end{eqnarray}


\vspace{6mm}
\noindent
{\large\bf Appendix C. Comments of John A. Dutton (1992)}
             \label{appendix_C}
\renewcommand{\theequation}{C.\arabic{equation}}
  \renewcommand{\thefigure}{C.\arabic{figure}}
   \renewcommand{\thetable}{C.\arabic{table}}
      \setcounter{equation}{0}
        \setcounter{figure}{0}
         \setcounter{table}{0}
\vspace{1mm}
\hrule
\vspace{3mm}

\noindent {\large {``Energetics with an entropy flavour''. Q. J. R. Meteorol. Soc. (1992), 118, pp. 165-166.}

J. A. Dutton acted as one of the two referees of my paper (the other one was R. P. Pearce).
The purpose of the note written by Dutton (1992) was to ``{\it provide an elementary motivation for, and a simple derivation of, the key results in the theory of available enthalpy presented by Marquet (1991)\/}''. 
The aim of Dutton (1992) was to ``{\it obtain a Bernoulli equation that governs the rates of change, following the motion, of energy quantities with entropy-like characteristics\/}''.

Dutton considered first that the specific enthalpy is equal to $h = c_p \: T$.
He then considered the same enthalpy equation (\ref{eq_8}) than in Marquet (1991):
\begin{equation}
dh/dt \; = \;  \alpha \: dp/dt  \: + \: q
\: ,
\nonumber
\end{equation}
where $q$ is the same rate of heating as in (\ref{eq_8}).

Then, Dutton ``{\it introduce entropy considerations by creating a covariance between the temperature $T$ and the quantity $q$\/}:
\begin{equation}
q \; = \; (T_r/T) \: q \: + \: (1-T_r/T) \: q
\: ,
\nonumber
\end{equation}
{\it where $T_r$ is a constant to be determined\/}''.

\begin{quote}
{\color{blue} My first remark is that it is indeed easy to do a job once the result is known!
The interesting feature of setting first $a_h=(h-h_r) - T_r \: (s-s_r) = c_p \: T_r \: {\cal F}(T/T_r -1)$ is that the equation for $da_h/dt$ given by (\ref{eq_9}) generate automatically the last term $(1-T_r/T) \: q$.
So why trying to reinvent first this Carnot's factor (efficiency), since it is a direct consequence of the exergy function $c_p \: T_r \: {\cal F}(T/T_r -1)$?
The use of the flow-exergy function $a_h$ avoid any uncertain trials and errors steps needed to discover (by chance) this term $(1-T_r/T) \: q$.
Moreover, both Carnot's factor and $a_h$ have been defined by founders of thermodynamics, long before publication of Dutton's papers (see Appendix~B).}
\end{quote}

Dutton interpreted the first term $(T_r/T) \: q$ by saying that, ``{\it for the specific entropy $s$\/}:,
\begin{equation}
\frac{T_r}{T} \: q 
\; = \; T_r \: \frac{q}{T}  
\; = \; T_r \: \frac{ds}{dt} 
\; = \; T_r \: \left[ \: 
    c_p \: \frac{d\ln T}{dt} 
  - R   \: \frac{d\ln p}{dt} 
            \: \right]
\: .
\nonumber
\end{equation}
He then mentioned that ``{\it we can define available energy forms \/}:
\begin{eqnarray}
a_T & = & \; h 
     \: - \: c_p \:  T_r 
     \: - \: c_p \:  T_r\: \ln(T/T_r)
\nonumber \\
a_p & = & R \:  T_r\: \ln(p/p_r)
\nonumber
\end{eqnarray}
{\it that vanish when $T=T_r=$constant and  $p=p_r=$constant.
The constants $T_r$ and $p_r$ can be determined as averages or specified as typical values\/}''.
(...) 
``{\it and thus we may define the available enthalpy as the sum of the components \/}:
\begin{equation}
a_h
\; = \; a_T \: + \: a_p
\; = \; (h - h_r) \:  - \: T_r \: (s - s_r)
\: .
\nonumber
\end{equation}

\begin{quote}
{\color{blue} What is new in this result, which is exactly the same as in (\ref{eq_3})? 
Dutton arrive at this definition of $a_h$ as a final stage of an improvised guessing game...
By doing this, he misses the important link with the general theory of (non-flow) exergy.
By doing this kind of guessing game, it is not possible to derive a moist-air generalization of $a_h$ and to arrive at the definition of the moist-air enthalpy $a_m$ published in (1993).
Really, the way the concept are defined are not all equivalent.
It is really important to realize that the concept of exergy corresponds to almost all the availability functions introduced in meteorology (except may be the one of McHall 1990-91), and that only exergy principles can be used as an entry point for moist-air generalizations.
}
\end{quote}

The last part of Dutton (1992) deals with the derivation of the Bernoulli equation valid  for $e_G + e_K + a_h$.
The conclusion of Dutton (1992) is that ``{\it we have succeeded in deriving a system of energy equations, following the motion, in which the thermodynamic quantities have an entropy flavour. 
There are, however, no implications that any conversions will actually occur or be of a known sign\/}''.

\begin{quote}
{\color{blue} Again, what is new in this result, which is exactly what I have derived in (\ref{eq_19}) and (\ref{eq_20})? 
This is just paraphrasing what I have explained in my paper.
I criticized implicitly in my paper the fact that Dutton (1973, 1976) has coined the quantity $T_0\:\Sigma$ a ``static entropic energy''.
I explained that $T_0\:\Sigma$ is exactly equal to what is called ``non-flow exergy'' in general Thermodynamics.
The aim of J. A. Dutton was probably to diminish the merit of my paper by showing that the quantity $a_h = a_T + a_p$ might be reinvent by using the same kind of heuristic arguments he has used in most of his papers and books.
Differently, I consider that atmospheric energetics is just a part of general thermodynamics, and since the concept of exergy exists in thermodynamics, we just need to use it, without any attempt to hide or reinvent this concept in the context of atmospheric energetics.
}
\end{quote}


\newpage
\vspace{5mm}
\noindent{\Large\bf References}
\vspace{2mm}

\noindent{$\bullet$ Bejan,~A.} {1987}.
{\it Advanced engineering thermodynamics.\/}
John Wiley \& Sons, New York.

\noindent{$\bullet$ Boer,~G.~J.} {1989}.
{On exact and approximate energy equations in pressure coordinates.}
{\it Tellus.\/}
{\bf 41A,} (2),
p.97--108.

\noindent{$\bullet$ Borel,~L.} {1987}.
{\it Thermodynamique et \'energ\'etique.\/}
Vol.~1.
Presses polytechniques Romandes, Lausanne.

\noindent{$\bullet$ Brennan,~F.~E. and Vincent,~D.~G.} {1980}.
{Zonal and eddy components of the synoptic-scale energy
budget during intensification of hurricane Carmen (1974).}
{\it Mon. Weather Rev.\/}
{\bf 108,}
p.954--965.

\noindent{$\bullet$ Carnot,~N,~L,~S.} {1824}.
{R\'eflexions sur la puissance motrice du feu, et sur les machines propres \`a d\'evelopper cette puissance.
See the account of Carnot's theory written by W. Thomson (1849) in the
{\it Trans. Roy. Soc. Edinb.\/}
{\bf 16},
p.541--574.
An English translation by R. H. Thurston of the version published in the ``Anales scientifique de l'\'Ecole Normale Sup\'erieure'' (ii. series, t.1, 1872) is available in the url: \url{http://www3.nd.edu/~powers/ame.20231/carnot1897.pdf} (Wiley \& Sons, 1897, digitized by Google)}

\noindent{$\bullet$ Clausius,~R.} {1865}.
{\"Uber verschiedene f\"ur die Anwendung bequeme Formen der
Hauptgleichungen der mechanischen W\"armetheorie.
(On Different Forms of the Fundamental Equations of the Mechanical Theory of Heat).
{\it Ann. der Phys. und Chem.\/}
{\bf 125},
p.353-400.}

\noindent{$\bullet$ Darrieus,~M.~G.} {1931}.
{L'\'evolution des centrales thermiques et la notion 
d'\'energie utilisable.}
{\it Science et Industrie,\/} {\bf 204}, 
p.122--126

\noindent{$\bullet$  Dutton,~J.~A. and Johnson,~D.~R.} {1967}.
{The theory of available potential energy and a variational
approach to atmospheric energetics.}
Pp.333--436, in Vol.~12 of 
{\it Advances in Geophysics.\/} 
Academic Press, New-York and London.

\noindent{$\bullet$ Dutton,~J.~A.} {1973}.
{The global thermodynamics of atmospheric motion.
{\it Tellus.\/}
{\bf 25,} (2),
p.89--110.}

\noindent{$\bullet$ Dutton,~J.~A.} {1976}.
{The ceaseless wind.
An introduction to the theory of atmospheric motion.
{\it McGraw-Hill.\/}
}

\noindent{$\bullet$ Dutton,~J.~A.} {1992}.
{Energetics with an entropy flavour.
{\it Q. J. R. Meteorol. Soc.\/}
{\bf 118},
p.165--166.} 

\noindent{$\bullet$ Evans,~R.~B.} {1969}.
{A Proof that Essergy is the only Consistent 
Measure of Potential Work.}
PhD Thesis, Dartmouth College, Hanover, New Hampshire.

\noindent{$\bullet$ Evans,~R.~B.} {1980}.
{Thermoeconomic isolation and essergy analysis.
{\it Energy.\/}
{\bf 5},
p.805--821.}

\noindent{$\bullet$ Feidt,~M.} {1987}.
{\it Thermodynamique et optimisation \'energ\'etique. 
Technique et documentation.\/}
Lavoisier, Paris.

 \noindent{$\bullet$ Gibbs,~J.~W.} {1873a}.
{Graphical methods in the thermodynamics of fluids.
{\it Trans. Connecticut Acad.\/}
{\bf II}: p.309--342.
(Pp 1--32 in Vol. 1 of
{\it The collected works of J. W. Gibbs,\/}
1928.
Longmans Green and Co.)} 

 \noindent{$\bullet$ Gibbs,~J.~W.} {1873b}.
{A method of geometrical representation
of the thermodynamic properties of substance
by means of surfaces.
{\it Trans. Connecticut Acad.\/}
{\bf II}: p.382--404.
(Pp 33--54 in Vol. 1 of
{\it The collected works of J. W. Gibbs,\/}
1928.
Longmans Green and Co.)} 

 \noindent{$\bullet$ Gibbs,~J.~W.} {1875-76-77-78}.
{On the equilibrium of heterogeneous substances.
{\it Trans. Connecticut Acad.}
{\bf III}: p.108--248, 1875-1876 and p.343-524, 1877-1878.
(Pp 55--353  in Vol. 1 of
{\it The collected works of J. W. Gibbs,\/}
1928.
Longmans Green and Co.)} 

\noindent{$\bullet$ Gill,~A.~E.} {1982}.
{\it Atmosphere-ocean dynamics.\/}
Academic Press, New York.

\noindent{$\bullet$ Glansdorff,~P. and Prigogine,~I.} {1971}.
{\it Structure stabilit\'e et fluctuations.\/}
{Masson Ed.} Paris
(Also available in English: {\it Thermodynamic theory of 
structure, stability and fluctuations\/}, Wiley-Interscience).

\noindent{$\bullet$ Gouy,~L.~G.} {1889}.
{Sur I'\'energie utilisable.
{\it Journal de physique th\'eorique et
appliqu\'ee.\/}
2\`eme s\'erie,
{\bf VIII},
p.501--518.}

\noindent{$\bullet$ Haywood,~R.~W.} {1974}.
{A critical review of the theorems of thermodynamic availability, with concise formulations. 
{\it J. Mech. Eng. Sci.}
  Part 1: Availability, 
{\bf 16}, (3), p.160--173;
  Part 2: Irreversibility, 
{\bf 16}, (4), p.258--267.}

\noindent{$\bullet$ Houghton,~J.~T.} {1977}.
{\it The physics of atmospheres.\/}
Cambridge University Press.

\noindent{$\bullet$  Johson,~D.~R. and Downey,~W.~K.} {1982}.
{On the energetics of open systems.
{\it Tellus.\/}
{\bf 34,} (5),
p.458--470.}

\noindent{$\bullet$  Jouguet,~E.} {1907}.
{Le th\'eor\`eme de M. Gouy et quelques-unes de ses applications.
{\it Revue de M\'ecanique.\/}
{\bf 20,} 
p.213--238.}

\noindent{$\bullet$ Karlson,~S.} {1982}.
{The exergy of incoherent electromagnetic radiation.
{\it Physica Scripta.\/}
{\bf 26},
p.329--33.}

\noindent{$\bullet$ Karlsson,~S.} {1990}.
{{\it Energy, Entropy and Exergy in the atmosphere.\/}
Thesis of the Institute of Physical Resource Theory.
Chalmers University of Technology.
 G\"oteborg, Sweden.}

\noindent{$\bullet$ Kasahara,~A.} {1974}.
{Various vertical coordinate systems used for numerical 
weather prediction.
{\it Mon. Weather Rev.\/}
{\bf 102},
p.509--522.}

\noindent{$\bullet$ Keenan,~J.~H.} {1932}.
{A steam chart for second-law analysis.
A study of Thermodynamic availability 
in the steam power plant.
{\it Mechanical Engineering\/}
{\bf 54},
p.195--204.}

\noindent{$\bullet$ Keenan,~J.~H.} {1951}.
{Availability and irreversibility in thermodynamics.
{\it Br. J. Appl. Phys.\/}
{\bf 2},
p.183--192.}

\noindent{$\bullet$ Kestin,~J.} {1980}.
{Availability: the concept and associated terminology.
{\it Energy.\/}
{\bf 5},
p.679--692.}

\noindent{$\bullet$ Koelher,~T.~L.} {1986}.
{A terrain-dependent reference atmosphere determination
method for available potential energy calculations.
{\it Tellus.\/}
{\bf 38A,} (1),
p.42--48.}

\noindent{$\bullet$ Kucharski,~F.} {1997}.
{On the concept of exergy and available potential energy.
{\it Q. J. R. Meteorol. Soc.\/}
{\bf 123},
p.2141--2156.} 

\noindent{$\bullet$ Landau,~L.,~D. and Lifchitz,~E.,~M.} {1958, 1976}.
{Statistical physics, (third edition). 
{\it Pergamon Press.}
London.}

\noindent{$\bullet$ Livezey,~R.~E. and Dutton,~J.~A.} {1976}.
{The entropic energy of geophysical fluid systems.
{\it Tellus.\/}
{\bf 28,} (2),
p.138--157.} 

\noindent{$\bullet$ Lorenz,~E.~N.} {1955}.
{Available potential energy and the 
 maintenance of the general circulation.
{\it Tellus.\/}
{\bf 7,} (2),
p.157--167.}

\noindent{$\bullet$ Lorenz,~E.~N.} {1967}.
{\it The nature and theory of the general 
      circulation of the atmosphere. \/}
W.M.O.

\noindent{$\bullet$ Lorenz,~E.~N.} {1978}.
{Available energy and the 
maintenance of a moist circulation.
{\it Tellus.\/}
{\bf 30}, (1),
p.15--31.}

\noindent{$\bullet$ Lorenz,~E.~N.} {1979}.
{Numerical evaluation of moist available energy.
{\it Tellus.\/}
{\bf 31}, (3),
p.230--235.}

\noindent{$\bullet$ McHall, ~Y.~L.}, {1990}.
{Available potential energy in the atmospheres.
{\it Meteorol. Atmos. Phys.\/}, 
{\bf 42}, 
39--55.}

\noindent{$\bullet$ McHall, ~Y.~L.}, {1991}.
{Available equivalent potential energy in moist atmospheres.
{\it Meteorol. Atmos. Phys.\/}, 
{\bf 45}, 
113--123.}

\noindent{$\bullet$ Marchal,~R.} {1956}.
{\it La thermodynamique et le th\'eor\`eme 
  de l'\'energie utilisable. \/}
Dunod, Paris.

\noindent{$\bullet$ Margules,~M.} {1901}.
{The mechanical equivalent of any given distribution of 
atmospheric pressure, and the maintenance of a given 
difference in pressure. 
{\it Smithsonian Miscellaneous collections.}
{\bf 51,} (4): 501--532, 1910
{\it (Translation by C. Abbe of a lecture read at the 
meeting of the Imperial Academy of Science, Vienna, 
July, 11, 1901, commemorating the Jubilee of the 
Central Institute for Meteorology and Terrestrial 
Magnetism)}.}

\noindent{$\bullet$ Margules,~M.} {1903-05}.
{On the energy of storms. 
{\it Smithsonian Miscellaneous collections,} 
51, 4, 533--595, 1910. 
{\it (Translation by C. Abbe from the appendix to the annual 
volume for 1903 of the Imperial Central Institute for Meteorology, 
Vienna, 1905.
`\"Uber die energie der st\"urme'. 
{\it Jahrb. Zentralantst. Meteorol.},
{\bf 40,} 1--26, 1903)}.}

\noindent{$\bullet$ Martinot-Lagarde,~A.} {1971}.
{\it Thermique classique et introduction \`a la m\'ecanique 
      des \'evolutions irr\'eversibles. \/}
Dunod, Paris.

\noindent{$\bullet$ Marquet~P.} {1991}.
{On the concept of exergy and available
enthalpy: application to atmospheric energetics.
{\it Q. J. R. Meteorol. Soc.}
{\bf 117}:
449--475.
}

\noindent{$\bullet$ Marquet~P.} {1993}.
{Exergy in meteorology: definition and properties
of moist available enthalpy.
{\it Q. J. R. Meteorol. Soc.}
{\bf 119} (511) :
567--590.} 

\noindent{$\bullet$ Marquet,~P.} {1995}.
{On the concept of pseudo-energy of T. G. Shepherd.
{\it Q. J. R. Meteorol. Soc.}
{\bf 121}:
455--459.}

\noindent{$\bullet$ Marquet,~P.} {2003a}.
{The available-enthalpy cycle. I: 
Introduction and basic equations.
{\it Q. J. R. Meteorol. Soc.}
{\bf 129}:
2445--2466.}

\noindent{$\bullet$ Marquet,~P.} {2003b}.
{The available-enthalpy cycle. II: 
Applications to idealized baroclinic waves.
{\it Q. J. R. Meteorol. Soc.}
{\bf 129}:
2467--2494.}

\noindent{$\bullet$ Maxwell,~J.~C.} {1871}.
{Theory of Heat. References are made in the text to next editions of this book. 
{\it Longmans, Green and Co.}
London.}

\noindent{$\bullet$ Michaelides,~S.~C.} {1987}.
{Limited area energetics of Genoa cyclogenesis.
{\it Mon. Weather Rev.\/}
{\bf 115},
p.13--26.}

\noindent{$\bullet$ Muench,~H.~S.} {1965}.
{On the dynamics of the wintertime stratosphere 
circulation.
{\it J. Atmos. Sci.\/}
{\bf 22},
p.349-360.}

\noindent{$\bullet$ Oort,~A.~H.} {1964}.
{On the energetics of the mean and eddy circulation 
in the lower stratosphere.
{\it Tellus.\/}
{\bf 16}, (3),
p.309--327.}

\noindent{$\bullet$ Oort,~A.~H., 
Ascher,~S.~C.,
Levitus,~S. and 
Peix\'oto,~J.~P.}
{1989}.
{New estimates of the available potential energy
  in the world ocean.
{\it J. Geophys. Res.\/}
{\bf 94}, (3),
p.3187--3200.}

\noindent{$\bullet$ Pearce,~R.~P.} {1978}.
{On the concept of available potential energy.
{\it Q. J. R. Meteorol. Soc.\/}
{\bf 104},
p.737--755.}

\noindent{$\bullet$ Pichler,~H.} {1977}.
{Die bilanzgleichung f\"ur die statischer 
entropische Energie der Atmos\-ph\"are.
{\it Arch. Met. Geoph. Biokl.\/}, 
Ser.A, 
{\bf 26},
p.341--347.}

\noindent{$\bullet$ Rant,~Z.} {1956}.
{Exergie, ein neues Wort f\"{u}r ``Technische 
 Arbeitsf\"{a}higkeit''.
{\it Forsch. Ing. Wes.\/}, 
{\bf 22},
p.36--37.}

\noindent{$\bullet$ Saltzman,~B. and Fleischer,~A.} {1960}.
{The modes of release of available potential energy 
  in the atmosphere.
{\it J. Geophys. Res.\/}
{\bf 65}, (4),
p.1215--1222.}

\noindent{$\bullet$ Smith,~P.~J., Vincent,~D.~D. and
Edmon,~H.~J.} {1977}.
{The time dependence of reference pressure in limited region
available potential energy budget equations.
{\it Tellus.\/}
{\bf 29}, (5),
p.476--480.}

\noindent{$\bullet$ Stodola,~A.} {1898}.
{Die Kreisprozesse der Gasmaschine.
{\it Zeit. V. D. I.\/}
{\bf 42}, (39),
p.1086--1091.}

\noindent{$\bullet$ Szargut,~J.} {1980}.
{International progress in second law analysis.
{\it Energy.\/}
{\bf 5}, (39),
p.709--718.}

\noindent{$\bullet$ Szargut,~J. and Styrylska,~T.} {1969}.
{Die exergetische Analyse von Prozessen der 
feuchten Luft (An exergetic analysis of 
processes for damp air).
{\it Heiz.-L\"uft.-Haustechn. (for: Heizung 
- L\"uftung - Haustechnik / Heating,  
Ventilation, Building services)}
{\bf 5},
p.173--178.}

\noindent{$\bullet$ Tait,~P.,~G.} {1879}.
{On the dissipation of energy. 
{\it Phil. Mag.\/}
{\bf 7}, 44, 5e series,
p.344--346.}

\noindent{$\bullet$ Taylor,~K.~E.} {1979}.
{Formulas for calculating available potential 
 energy over uneven topography.
{\it Tellus.\/}
{\bf 31}, (3),
p.236--245.}

\noindent{$\bullet$ Thomson,~W.} {1849}.
{An account of Carnot's theory of the ``Motive Power of Heat'',
 with numerical results deduced from Regnault's experiments on 
steam. 
{\it Trans. Roy. Soc. Edinb.\/}
{\bf 16,} Part~5,
p.541--574.}

\noindent{$\bullet$ Thomson,~W.} {1853}.
{On the restoration of mechanical energy 
from an unequally heated space.
{\it Phil. Mag.\/}
{\bf 5,} 30, 4e series,
p.102--105.}

\noindent{$\bullet$ Thomson,~W.} {1879}.
{On thermodynamic motivity. 
{\it Phil. Mag.\/}
{\bf 7,} 44, 5e series,
p.346--352.} 

\noindent{$\bullet$ Van Mieghem,~J.} {1956}.
{The energy available in the atmosphere for 
  conversion into kinetic energy. 
{\it Beitr. Phys. Armos.\/}
{\bf 29,}
p.129--142.}

\end{document}